\documentclass[review]{elsarticle}

\usepackage{lineno, hyperref, siunitx}
\usepackage{amssymb,amsmath}
\usepackage[margin=1in]{geometry}
\usepackage{textpos}

\usepackage{enumitem}
\usepackage[parfill]{parskip}    
\usepackage{latexsym, amssymb, bbm}
\usepackage{amsmath}
\usepackage{amsfonts}
\usepackage{amsthm}

\usepackage{siunitx}
\usepackage{verbatim}
\usepackage{fancyhdr}
\usepackage{mathrsfs}
\usepackage{enumerate}
\usepackage{geometry}
\usepackage{graphicx}
\usepackage{algorithm}
\usepackage{algpseudocode}
\usepackage{caption}
\usepackage{subcaption}
\usepackage{cancel}
\usepackage{listings}
\usepackage{xcolor}
\usepackage{booktabs}
\usepackage{listings}
\modulolinenumbers[5]

\journal{arXiv}

\begin{document}

\begin{frontmatter}

\title{Calibrating a Finite-strain Phase-field Model of Fracture for Bonded Granular Materials with Uncertainty Quantification}


\author[cvenaddress]{Abigail C.~Schmid}
\author[cvenaddress]{Erik Jensen\corref{erik}}
\author[csaddress]{Fabio Di Gioacchino}
\author[utdmechanicaladdress]{Pooyan B.~Javadzadeh}
\author[lanlsigma]{Nate E.~Peterson}
\author[cvenaddress]{C.~Gus Becker}
\author[utdmechanicaladdress]{Hongbing Lu}
\author[cvenaddress,appmaddress]{Fatemeh Pourahmadian}
\author[lanlsigma,minesaddress]{Amy J.~Clarke}
\author[asenaddress,appmaddress]{Alireza Doostan}
\author[cvenaddress]{Richard A.~Regueiro}
\cortext[erik]{Erik.Jensen@colorado.edu}

\address[cvenaddress]{Department of Civil, Environmental, and Architectural Engineering, University of Colorado Boulder. 1111 Engineering Dr, Boulder, CO 80309 USA}
\address[csaddress]{Department of Computer Science, University of Colorado Boulder. 1111 Engineering Dr, Boulder, CO 80309 USA}
\address[utdmechanicaladdress]{Department of Mechanical Engineering, University of Texas at Dallas. 800 W. Campbell Road Richardson, TX 75080 USA}
\address[lanlsigma]{Sigma Manufacturing Science Division, Los Alamos National Laboratory. P.O. Box 1663, Los Alamos, NM 87545 USA}
\address[minesaddress]{George S. Ansell Department of Metallurgical and Materials Engineering, Colorado School of Mines. 1301 19th St, Hill Hall Golden, CO 80401 USA}
\address[asenaddress]{Ann \& H.J. Smead Department of Aerospace Engineering Sciences, University of Colorado Boulder. 3775 Discovery Dr, Boulder, CO 80303 USA}
\address[appmaddress]{Department of Applied Mathematics, University of Colorado Boulder. 11 Engineering Dr, Boulder, CO 80309 USA}

\begin{abstract}

To study the mechanical behavior of mock high explosives, an experimental and simulation program was developed to calibrate, with quantified uncertainty, a material model of the bonded granular material Idoxuridine and nitroplasticized Estane-5703. This paper reports on the efficacy of such a framework as a generalizable methodology for calibrating material models against experimental data with uncertainty quantification. Additionally, this paper studies the effect of two manufacturing temperatures and three initial granular configurations on the unconfined compressive behavior of the resulting bonded granular materials. In each of these cases, the same calibration framework was used; in that, hundreds of high-fidelity direct numerical simulations using a new, GPU-enabled, high-performance finite element method software, Ratel, were run to calibrate a finite-strain phase-field fracture model against experimental data. It was found that manufacturing temperature influenced the elastic response of the mock high explosives, with higher temperatures yielding a stiffer response. By contrast, it was found that the initial configuration of the grains had a negligible impact on the overall behavior of the mock high explosives, though it remains possible that local damage accumulation within the specimens could be altered by the initial configurations. Overall, the calibration framework was successful at creating well-calibrated models, showing its usefulness as an engineering and scientific tool. 
    
\end{abstract}


\end{frontmatter}


\section{Introduction}
\label{sec:intro}

    The study of mechanical mock high explosives is of significant interest from the perspectives of both general scientific inquiry and national security and defense initiatives. While non-energetic, studying mechanical mocks allows insights into the mechanical behavior of bonded granular materials, such as how they deform under loading. This mechanical behavior is the result of the underlying microstructure where grains of mock explosives are bonded together by a polymer binder \cite{YeagerNanoindentation2012}. Grain-scale mechanics and crack initiation and propagation inform the behavior of the composite.

    In this work, a mock high-explosive composite of Idoxuridine (IDOX) and nitroplasticized Estane-5703 (Estane) was studied. This is a newer mock and has been the focus of several experimental studies in recent years \cite{CadyMechanical2006, YeagerDevelopment2018, YeagerDevelopment2019, BurchNanoindentation2017, BurchCompressive2022, BowdenSolubilityComparison2020, LiuPBX95012020, LiuBrazilian2023, HermanComposite2021, SchmidEnsemble2024}. The goal was to study the influence of the manufacturing process and the effects of the initial microstructure on the mechanical behavior of the resulting composite. To do this, many specimens of the composite were manufactured by pressing carefully measured amounts of the two constituents in a cylindrical die with a load frame heated to two different temperatures. Three of those specimens were imaged using computed tomography (CT) to provide images of the microstructure, and twelve of the specimens were used in unconfined compression experiments to capture one example of the possible behavior of the composite via the experimental force vs.~displacement response. 
    
    From the CT images, finite element direct numerical simulations (DNS) were developed by segmenting the images and generating three 3D meshes of the initial geometric conditions. Finite-strain phase-field damage constitutive models were used to capture the possible fracture of both the IDOX and Estane constituents observed during the experiments. To study the influence of different manufacturing temperatures and three initial geometries, the constitutive model parameters were calibrated with quantified uncertainty using the experimental data. The uncertainty quantification framework is in the form of Bayesian inference, where the posterior distribution provides a range of plausible values for the model parameters. This approach is similar to recent calibration work with equations of state (EOS) for the high-explosives this mock is designed to mimic \cite{LeidingPBX90122023, LindquistUncertainty2023, SchmidPosterior2024, AndrewsCalibration2024, LindquistSensitivity2025}. However, the types of model parameters are different, as the EOSs are often a functional form with adjustable parameters that can be solved rapidly, whereas the constitutive model parameters calibrated here are embedded in a finite element simulation. 
    
\begin{figure}[h!]  
        \centering
        \includegraphics[width=0.9\textwidth]{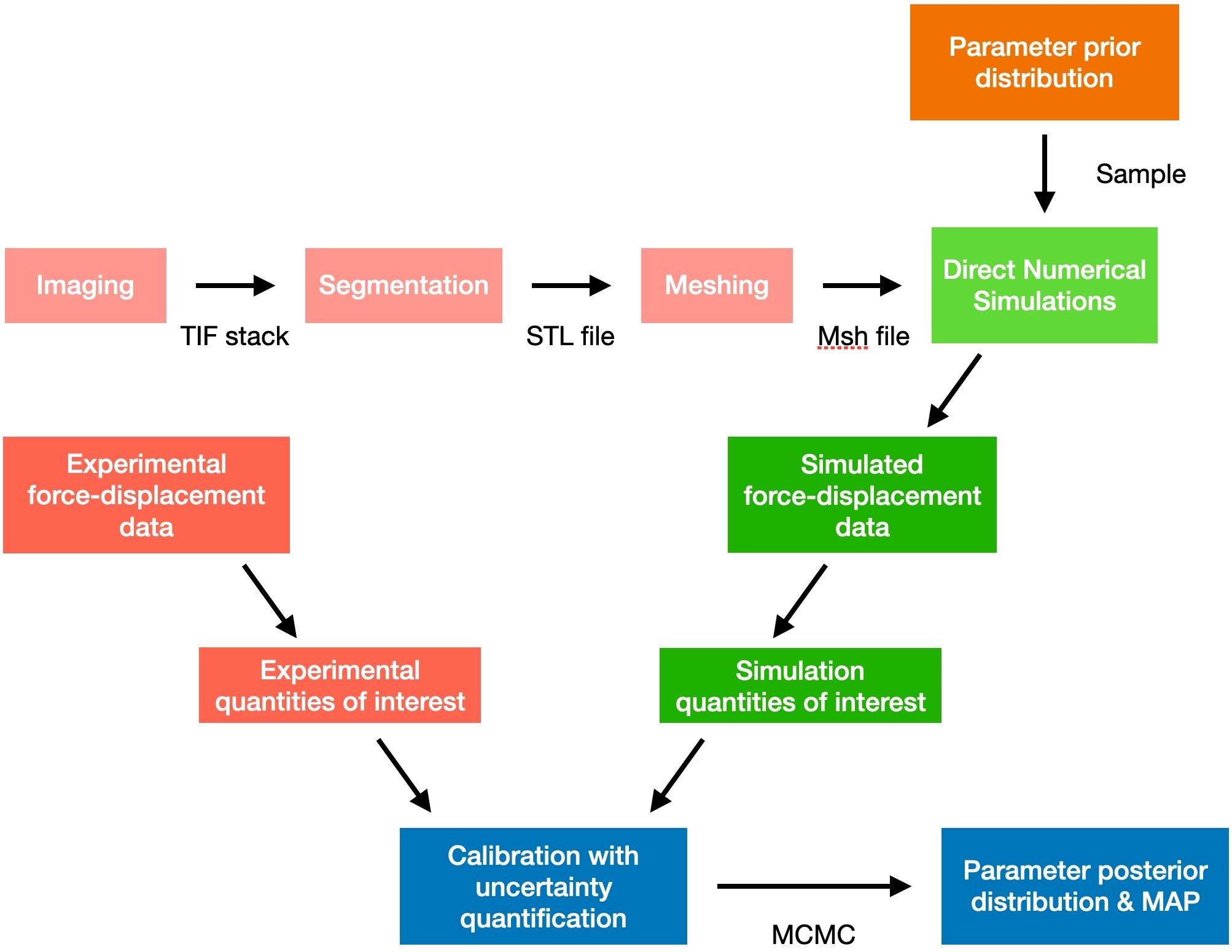}
        \caption{Analysis workflow.}
        \label{fig:workflow}
    \end{figure}

    Hundreds of simulations were run using a supercomputer with material properties sampled from initial guess (prior) distributions. Then, the simulation results were used to train a surrogate model in the form of a polynomial chaos expansion \cite{WienerHomogeneous1938, XiuWeiner2002, XiuModeling2003, GhanemStochastic2003, DoostanNon2011, HamptonCompressive2015, SoizePhysical2004, ErnstConvergence2012, OHaganPolynomial2013} to speed up computations in the Bayesian inverse problem. Finally, the posterior distributions were computed using Markov Chain Monte Carlo \cite{PyMC2023}, and the resulting distributions over the material properties were compared to study the manufacturing influences. In summary, an overview of the workflow approach is presented in Figure \ref{fig:workflow}. Two concurrent prongs of the workflow see quantities of interest pulled from experimental and DNS data compared against one another to calibrate posterior distributions for the varied material properties in the simulations. From there, the posterior distribution could be sampled, or a single point estimate (maximum \textit{a posteriori}) could be used to run calibrated behavior simulations for further investigations. 

    The purpose of this paper is to report on the efficacy of this calibration framework both as a method of analyzing the effects of processing and initial conditions on the ultimate behavior of materials such as mock high explosives and also as a general calibration workflow useful in many other engineering and scientific contexts. Formally quantifying the uncertainty of a calibration via posterior distributions can greatly improve one's understanding of a system as compared to a single set of calibrated values. The rest of this article is organized as follows. Section \ref{sec:experiments} explains the experimental methods and data. Section \ref{sec:dns} describes the details of direct numerical simulations. Section \ref{sec:workflow} summarizes the overall workflow. Section \ref{sec:uq} outlines the Bayesian inference framework used for the calibration with uncertainty quantification. Section \ref{sec:results} highlights the results of the calibrations. Lastly, Section \ref{sec:conclusions} summarizes the conclusions and directions for future work. 

\section{Experimental Campaign and Data}
\label{sec:experiments}

\subsection{Quasi-static Unconfined Compression}
\label{sec:compression}

\begin{figure}[!h]
        \centering
        \includegraphics[scale=0.5]{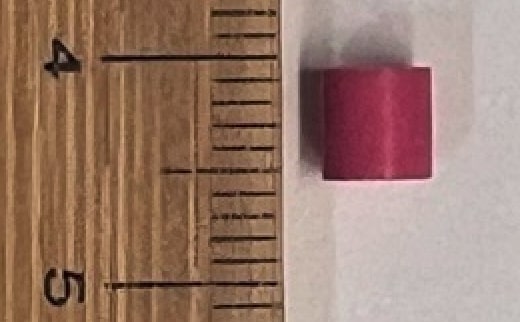}
        \caption{One of the IDOX/Estane specimens.}
        \label{fig:puck}
    \end{figure}

    The experimental data used to calibrate the material models were force-displacement curves measured from quasi-static unconfined compression tests. The composite specimens of IDOX and nitroplasticized Estane were made of IDOX (95\% by weight), BDNPA/F (2.5\% by weight), and Estane 5703 (2.5\% by weight). The IDOX particles had a bimodal size distribution where half of the particles were less than 75 \si{\micro\meter} in size and half were larger than 150 \si{\micro\meter}. The specimens were manufactured by pressing the aforementioned mixture in confined compression to $3,000\ \si{\newton}$ using a load frame heated to either $50\ \si{\degreeCelsius}$ or $90\ \si{\degreeCelsius}$ \cite{CamerloPressing2021}. Figure \ref{fig:puck} shows what the manufactured specimens looked like. Table \ref{tab:exp-values} lists the geometry of each specimen. The specimens had a radius of $2.5\ \si{\milli\meter}$ and a height of approximately $5\ \si{\milli\meter}$. The height variance was due to minor variations in the manufacturing process and the force-controlled pressing operation. Ten specimens were manufactured at $50\ \si{\degreeCelsius}$ and four at $90\ \si{\degreeCelsius}$. 

\begin{figure}[!h]
        \centering
        \subfloat[Initial]{\includegraphics[width=.32\textwidth]{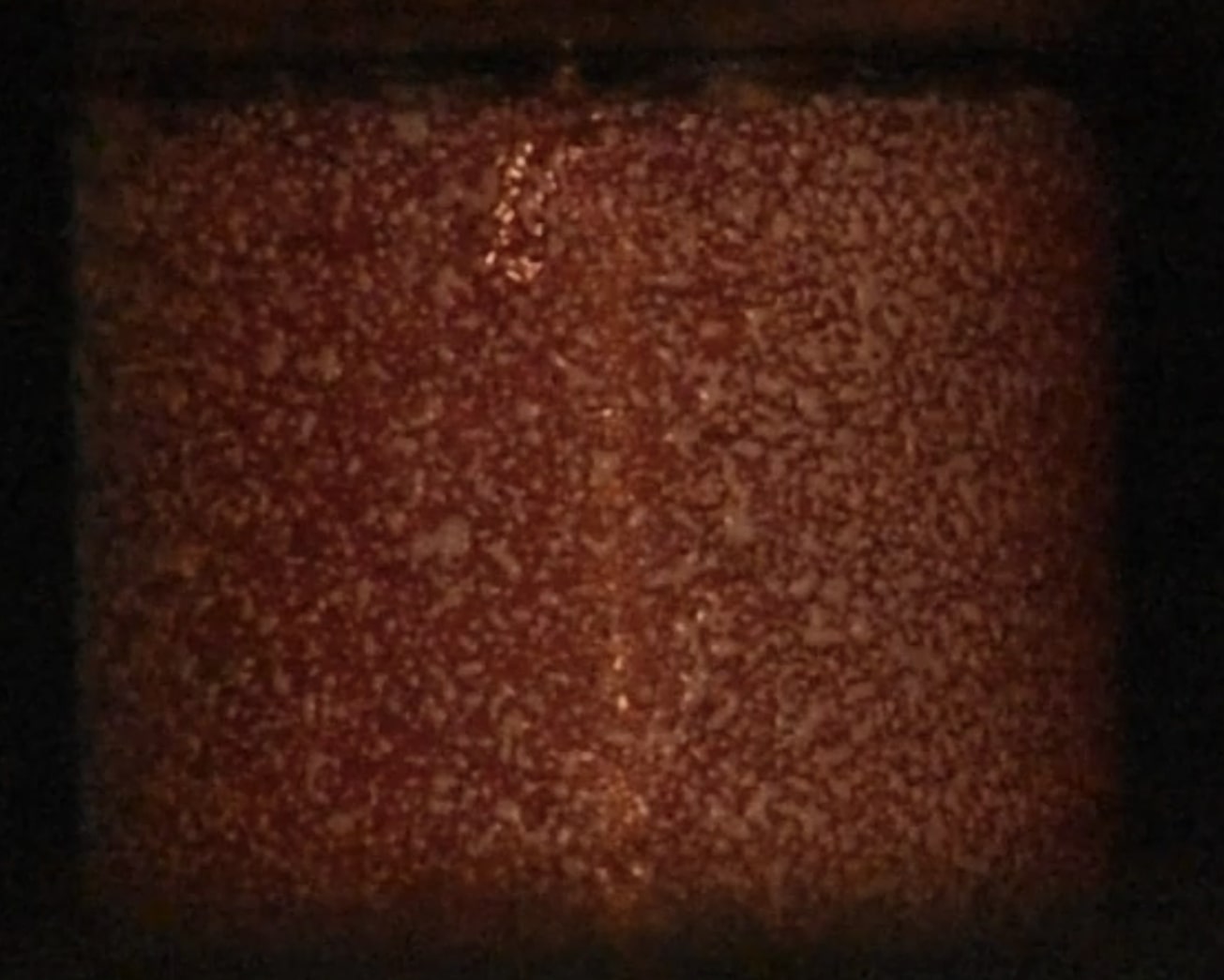}}
        \subfloat[Peak force]{\includegraphics[width=.32\textwidth]{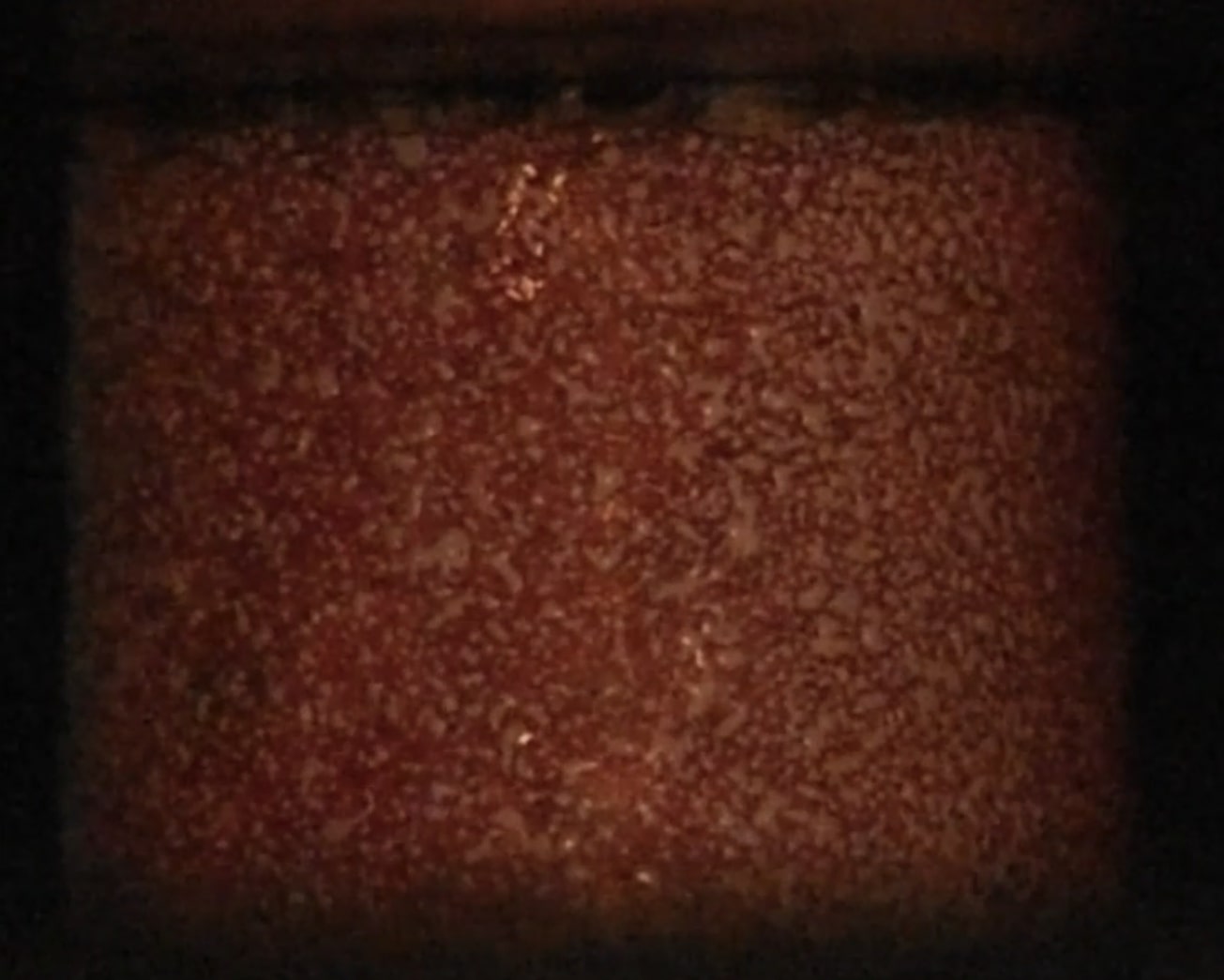}}
        \subfloat[Failure]{\includegraphics[width=.32\textwidth]{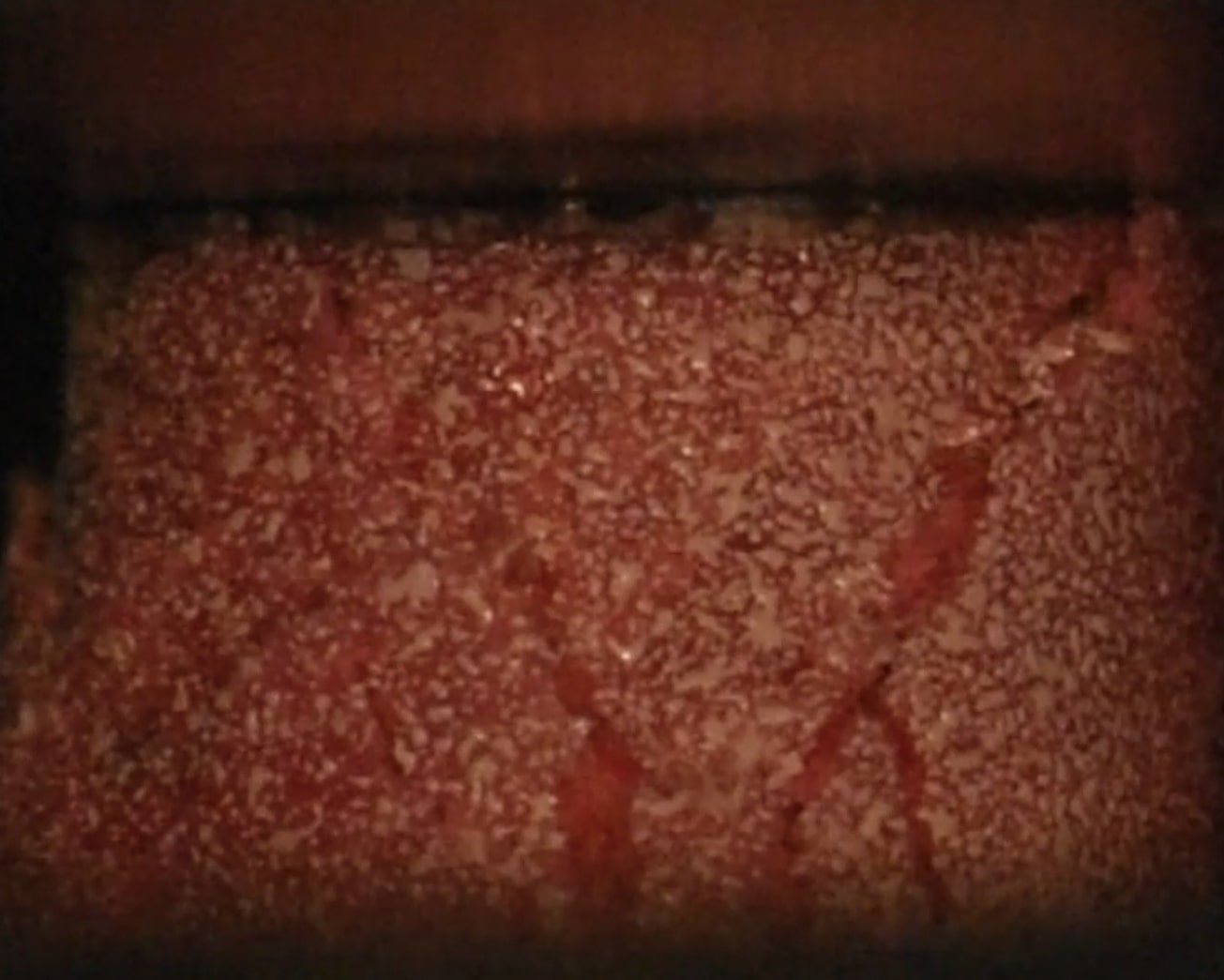}}
        \caption{Snapshots of the quasi-static compression test on sample S-50-01.}
        \label{fig:exp-snapshots}
    \end{figure}

\begin{figure}[!h]
        \centering
        \includegraphics[width=1.0\textwidth]{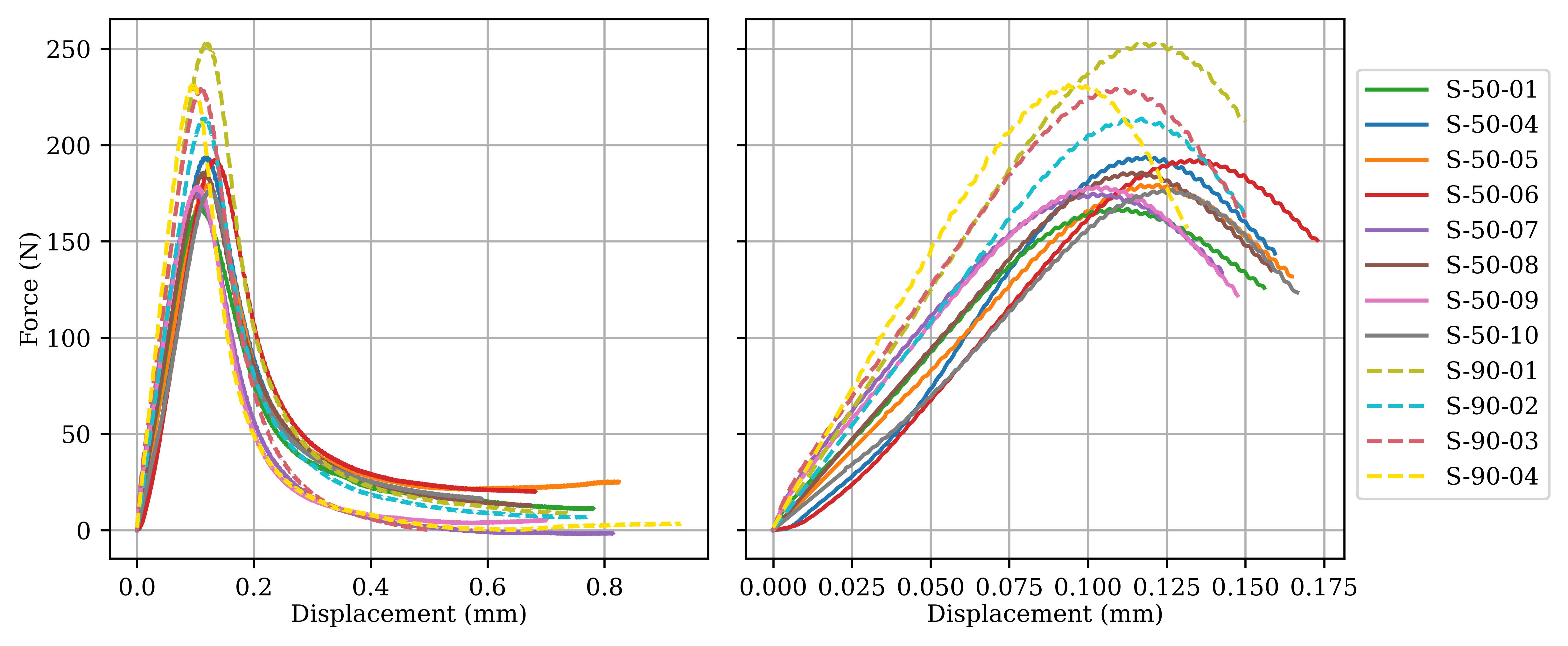}
        \caption{Experimental force-displacement curves for specimens manufactured at $50\ \si{\degreeCelsius}$ (solid lines) and $90\ \si{\degreeCelsius}$ (dashed lines). Left: full experimental measurements. Right: a truncated view of the data which matches the range of the simulations. This truncated data was used to compute the experimental quantities of interest. }
        \label{fig:all-exp-data}
    \end{figure}

    For eight of the specimens manufactured at $50\ \si{\degreeCelsius}$ and all four $90\ \si{\degreeCelsius}$ specimens, unconfined compression tests were conducted at ambient temperature with an Instron 5969 materials testing machine equipped with a 500 N Instron load cell (Catalog\# 2580-105) at the University of Texas at Dallas. Each specimen was placed at the center of $\SI[number-unit-product=\text{-}]{2}{inch}$ platens, and petroleum jelly (Vaseline) was applied to both ends of the specimen to reduce surface friction. The specimens were compressed to failure from above (approximately $0.6-0.8\ \si{\milli\meter}$ of displacement) at a strain rate of $1.7 \times 10^{-3}\ \si{1\per\second}$ following ASTM D695 \cite{ASTM_D695}. Figure \ref{fig:exp-snapshots} shows snapshots of the S-50-01 sample in compression at three points during the test. Figure \ref{fig:all-exp-data} shows the force-displacement data collected during the tests and used in this study. 

    \begin{table}[h!]
    \centering
    \begin{tabular}{lcccl}
        \hline
        Sample ID & \begin{tabular}{@{}c@{}} Sample height \\ ($\si{\milli\meter}$) \end{tabular} & \begin{tabular}{@{}c@{}} Sample radius \\ ($\si{\milli\meter}$) \end{tabular} & \begin{tabular}{@{}c@{}} Sample volume \\ ($\si{\milli\meter\cubed}$) \end{tabular} & Use \\
        \hline
        \hline
        S-50-01 & 4.95 & 2.50 & 97.20 & CT \& FD \\
        S-50-02 & 4.67 & 2.50 & 91.73 & CT \\ 
        S-50-03 & 5.11 & 2.50 & 100.19 & CT \\ 
        S-50-04 & 5.00 & 2.50 & 98.20 & FD \\ 
        S-50-05 & 5.00 & 2.50 & 98.20 & FD \\ 
        S-50-06 & 5.00 & 2.50 & 98.20 & FD \\ 
        S-50-07 & 5.00 & 2.50 & 98.20 & FD \\ 
        S-50-08 & 5.03 & 2.50 & 98.69 & FD \\ 
        S-50-09 & 5.06 & 2.50 & 99.20 & FD \\ 
        S-50-10 & 4.98 & 2.50 & 97.69 & FD \\ 
        S-90-01 & 5.21 & 2.50 & 102.19 & FD \\ 
        S-90-02 & 5.33 & 2.50 & 104.68 & FD \\ 
        S-90-03 & 5.41 & 2.50 & 106.17 & FD \\ 
        S-90-04 & 5.00 & 2.50 & 98.20 & FD \\ 
        \hline
    \end{tabular}
    \caption{Geometry of the experimentally tested specimens. The last column indicates if the sample was tested via quasi-static compression to generate force-displacement data = FD or scanned with CT = CT. }
    \label{tab:exp-values}
    \end{table}

\subsection{Computed Tomography}
\label{sec:CT}

    Computed tomography (CT) scans were used to generate the initial geometry for the simulations (see Section \ref{sec:meshing}). The CT data were collected for three specimens using a Zeiss Xradia 520 Versa X-ray $\mu$CT  machine (Zeiss, Baden-Württemberg, Germany). Before conducting the CT scans, the force-displacement response from quasi-static compression was carefully analyzed to determine suitable pause points for in-situ imaging. This approach enabled the capture of the microstructural evolution both prior to and after the peak load, which facilitated a detailed examination of the grain behavior throughout the deformation process. Three scans were performed in total, requiring approximately eight hours each, which resulted in a cumulative scan time of roughly 24 hours. Note that the S-50-01 sample was used for both CT and quasi-static compression testing. As a result of having both data types, this sample was used throughout this workflow for development and testing purposes. 
    
\begin{figure}[!h]
        \centering
        \subfloat[Raw CT image (false color)]{\includegraphics[width=.32\textwidth]{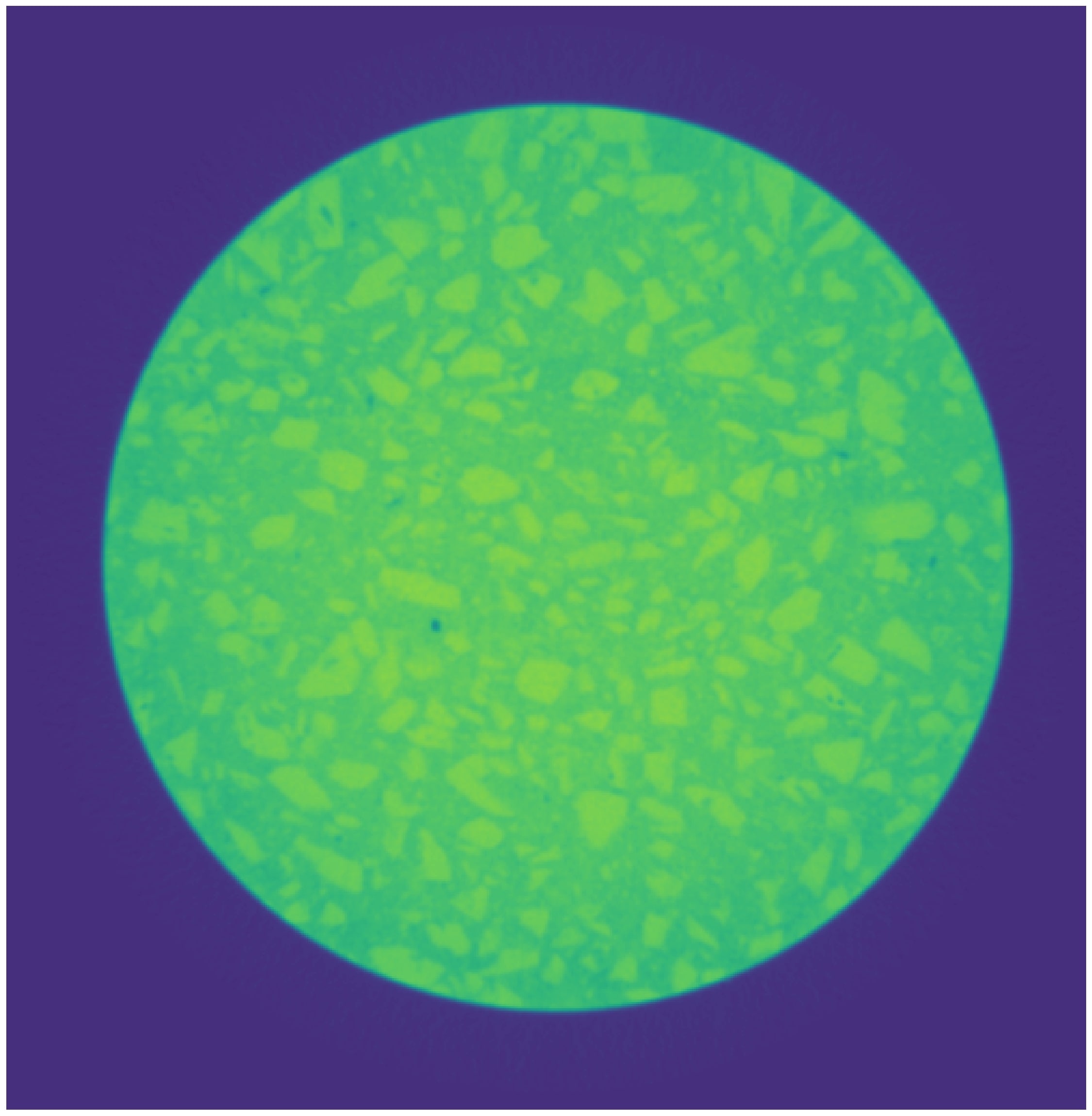}}
        \subfloat[Semantic Segmentation]{\includegraphics[width=.32\textwidth]{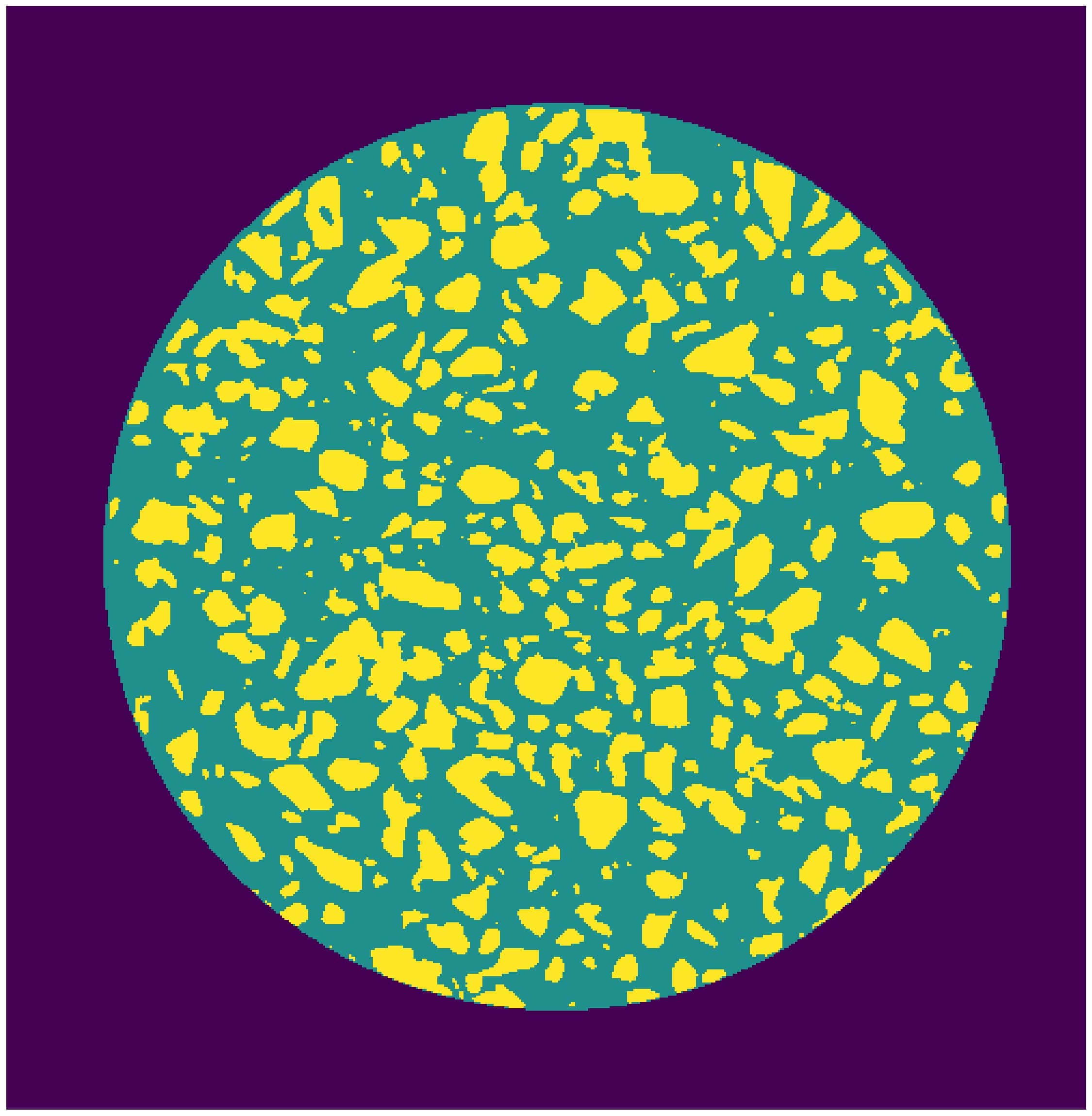}}
        \subfloat[Instance Segmentation]{\includegraphics[width=.32\textwidth]{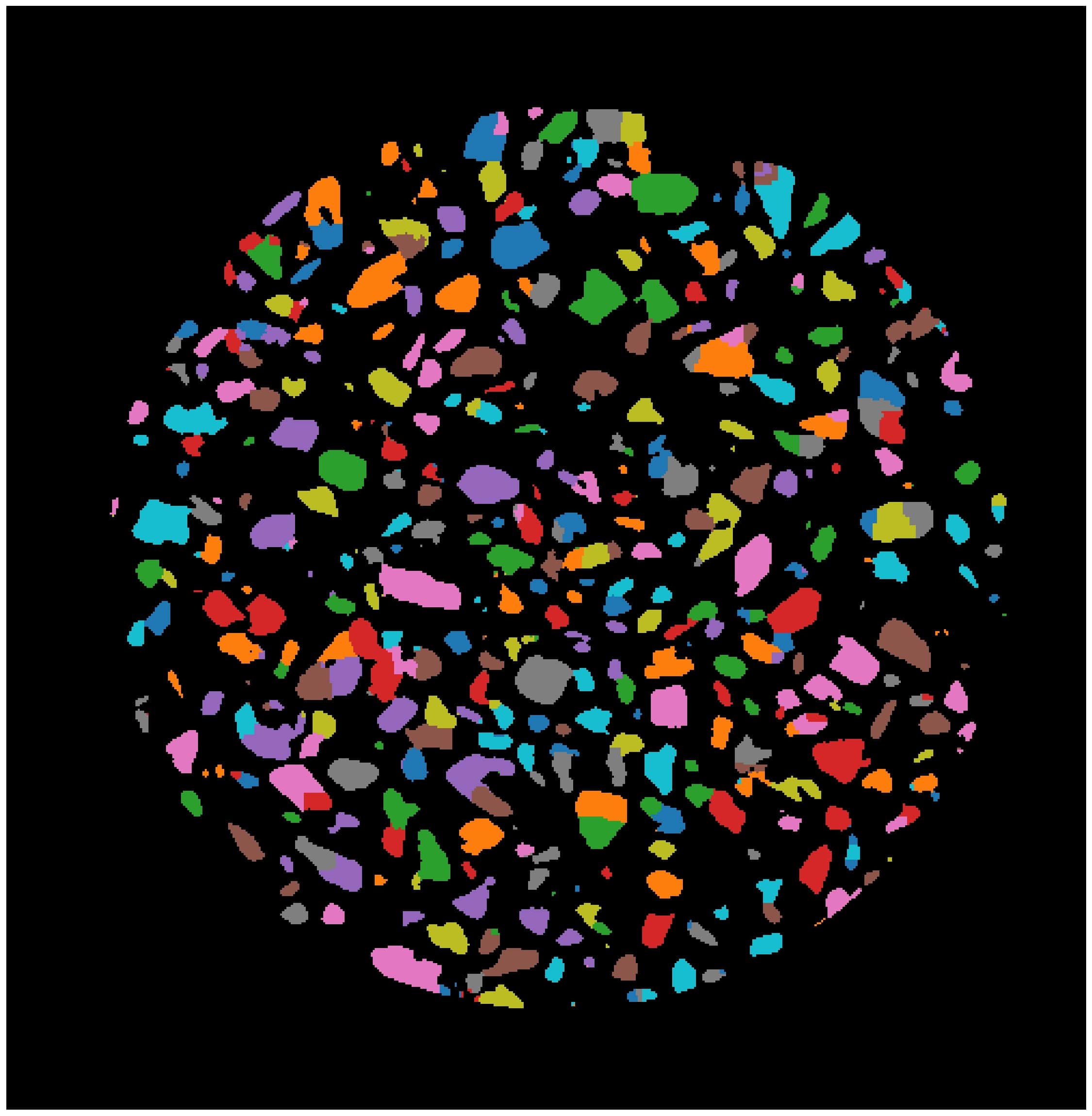}}
        \caption{Steps in the segmentation process shown by cross-sectional area slices taken through the composite sample at roughly mid-height: (a) raw imaging data (false color), (b) Semantic segmentation distinguishing between the IDOX grains in yellow and the Estane binder in blue, and (c) instance segmentation resulting from the application of a watershed algorithm, differentiating each grain by color.}
        \label{fig:segmentflow}
    \end{figure}

    Figure \ref{fig:segmentflow} shows three cross-sections through the S-50-01 sample. The first panel (a) is CT data collected immediately post-sample manufacturing, rendered in false color to highlight intensity differences between the IDOX grains and the Estane binder. To prepare the initial geometry for the unconfined compression simulations, an image processing routine was applied to the sample in 3D using the Python packages Segmentflow \cite{segmentflow} and scikit-image \cite{scikit-image}. The second panel (b) shows the semantic segmentation after a manual threshold was applied to distinguish the binder from void and the grains from the binder. The third panel (c) shows the instance segmentation of the grains after separation from the binder. This was the result of a watershed algorithm, as implemented in scikit-image. Note that there exists some degree of over-segmentation, which is due to the large range of grain sizes, and the difficulty in setting a minimum distance to filter the segmentation seeds such that small particles would not be lost. Additional details are provided in Section \ref{sec:meshing}.

\subsection{Scanning Electron Microscopy}
\label{sec:sem}

    In the CT scans and resulting meshes (Section \ref{sec:meshing}), the grain volume fraction was found to be small compared to the value expected based on the manufacturing procedure. The specimens were manufactured to a mass ratio of IDOX to Estane of 19:1, and the resulting volume fraction of IDOX grains was expected to be approximately 0.92 in the composite \cite{CamerloPressing2021}. However, as is seen in Table \ref{tab:mesh-values}, the finite element meshes had grain fractions between 0.19 and 0.27. In the underlying microstructure of the specimens, shown in Figure \ref{fig:segmentflow} (middle), the yellow patches in the middle image are large particles of IDOX. From visual inspection, volume fractions near 0.25 seem reasonable. To find the missing IDOX, scanning electron microscopy (SEM) imaging was performed to look at scales smaller than what was visible in the CT scans. 
    
    Specimens of the mock were prepared for SEM using a similar procedure to that outlined in \cite{SkidmoreMicroscopial1997}. Following the final polishing step ($0.05\ \si{\micro\meter}$ colloidal silica), a thin layer of gold was applied to the specimens using a sputter coater to make the specimens electrically conductive. SEM imaging (SE/BSE) was performed using a Tescan S8252G SEM/FIB instrument at the Colorado School of Mines. The images were collected with the microscope accelerating voltage and beam current set to $10\ \si{\kilo\volt}$ and $1\ \si{\nano \ampere}$, respectively. 

    \begin{figure}[!h]
        \centering
        \includegraphics[width=1.0\textwidth]{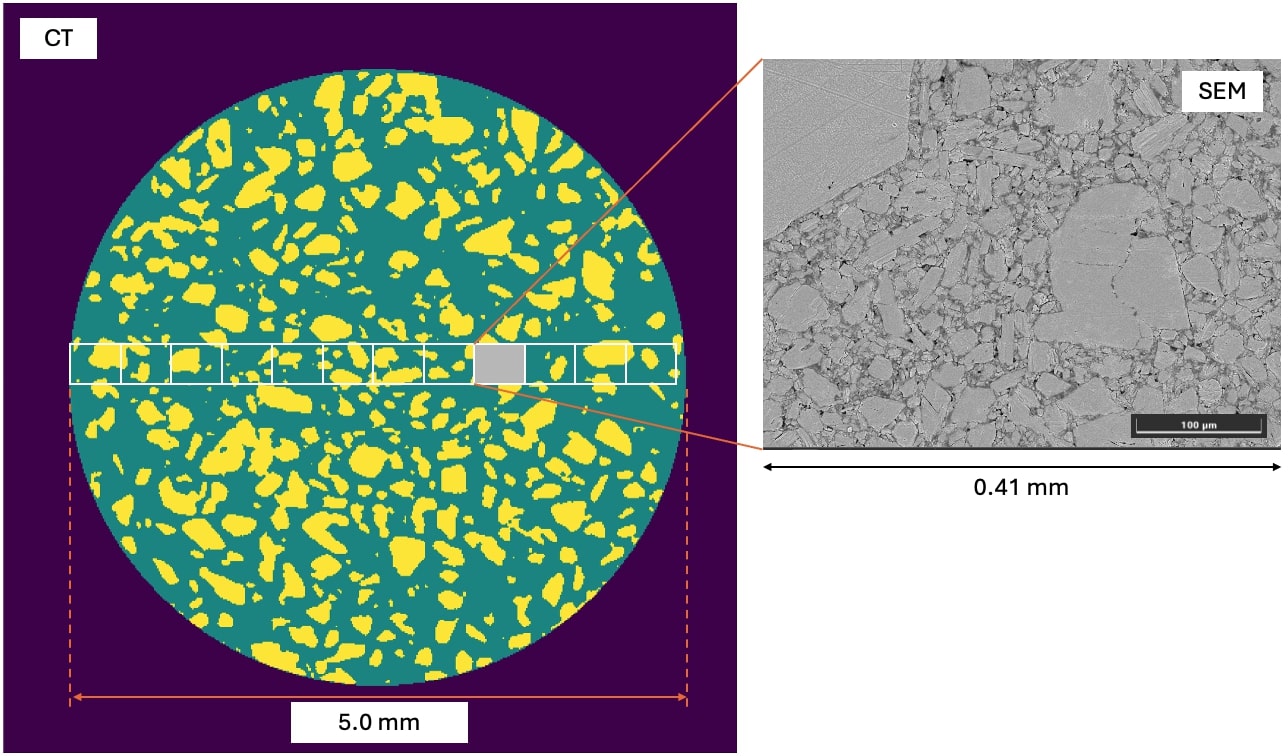}
        \caption{Comparison between segmented CT and SEM images for IDOX/Estane composites. Note that the CT scan was taken on S-50-01 and the SEM was taken on S-50-02, but both specimens were made using the same materials and procedures.}
        \label{fig:SEM-CT-comparison}
    \end{figure}    
    
    An SEM image from the S-50-02 sample is shown in Figure \ref{fig:SEM-CT-comparison} in comparison to the middle figure from Figure \ref{fig:segmentflow}, and it illuminates that between the larger IDOX grains that are clearly visible in the CT, there were smaller IDOX fragments mixed in with the Estane that were too small to discern at CT resolutions (the teal areas). For this study, the mixture of small IDOX pieces and the Estane binder is referred to as the ``matrix'', and it was modeled as a single, homogeneous material. It was expected that the matrix would be significantly stiffer than the Estane binder alone because of the inclusion of small IDOX pieces; however, it was unclear how much stiffer or what other properties might change compared to experimentally determined values for IDOX \cite{YeagerDevelopment2019, BurchNanoindentation2017} and Estane \cite{CadyMechanical2006, Estane} individually. From test simulations, it was found that an appropriate Young's modulus for the matrix is approximately $250\ \si{\mega\pascal}$, which is three orders of magnitude larger than a reasonable value for Estane ($0.46\ \si{\mega\pascal}$ \cite{CadyMechanical2006}). The calibration in this work focused on determining appropriate values for the matrix. 


\section{Finite Element Simulations}
\label{sec:dns}

\subsection{Finite Element Meshing}
\label{sec:meshing}

    A segmentation and surface meshing workflow was created using the Python package Segmentflow \cite{segmentflow} to segment CT scans of manufactured specimens and generate a collection of 3D surface meshes representing the IDOX grains. The Segmentflow workflow first loads the raw CT image data and pre-processes the raw images to sharpen the intensity differences between matrix and grain. The pre-processed images are separated into void, matrix, and grain classes using user-supplied thresholds at different intensity levels/densities (semantic segmentation, see Figure \ref{fig:segmentflow} (b)). The voxels (volume pixels) corresponding to the grain class are subject to a watershed segmentation to differentiate between different IDOX grains (Figure \ref{fig:segmentflow} (c) with each color now representing a single IDOX grain). The segmented volumes were used to generate a collection of surface meshes using a marching cubes algorithm \cite{MarchingCubes1987} to generate Standard Triangle Language (STL) files of each IDOX grain, then the STLs were post-processed by smoothing the surfaces to make them less blocky and decimation to reduce the complexity. The entire procedure is described in more detail in \cite{AppletonCT2025}.
 
\begin{figure}[!h]
        \centering
        \subfloat[CT resolution grain surface mesh]{\includegraphics[width=.32\textwidth]{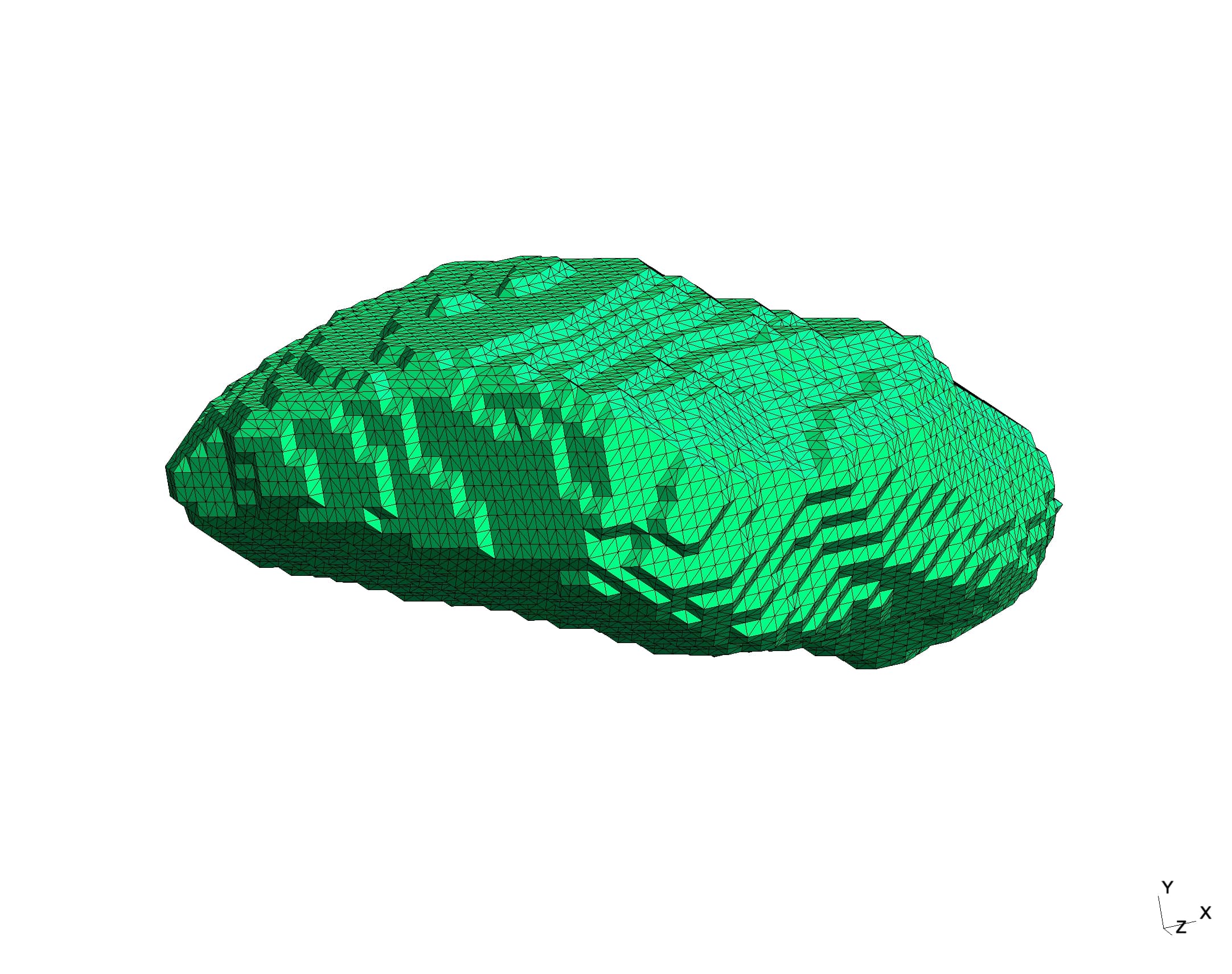}}
        \subfloat[Smoothed grain surface mesh]{\includegraphics[width=.32\textwidth]{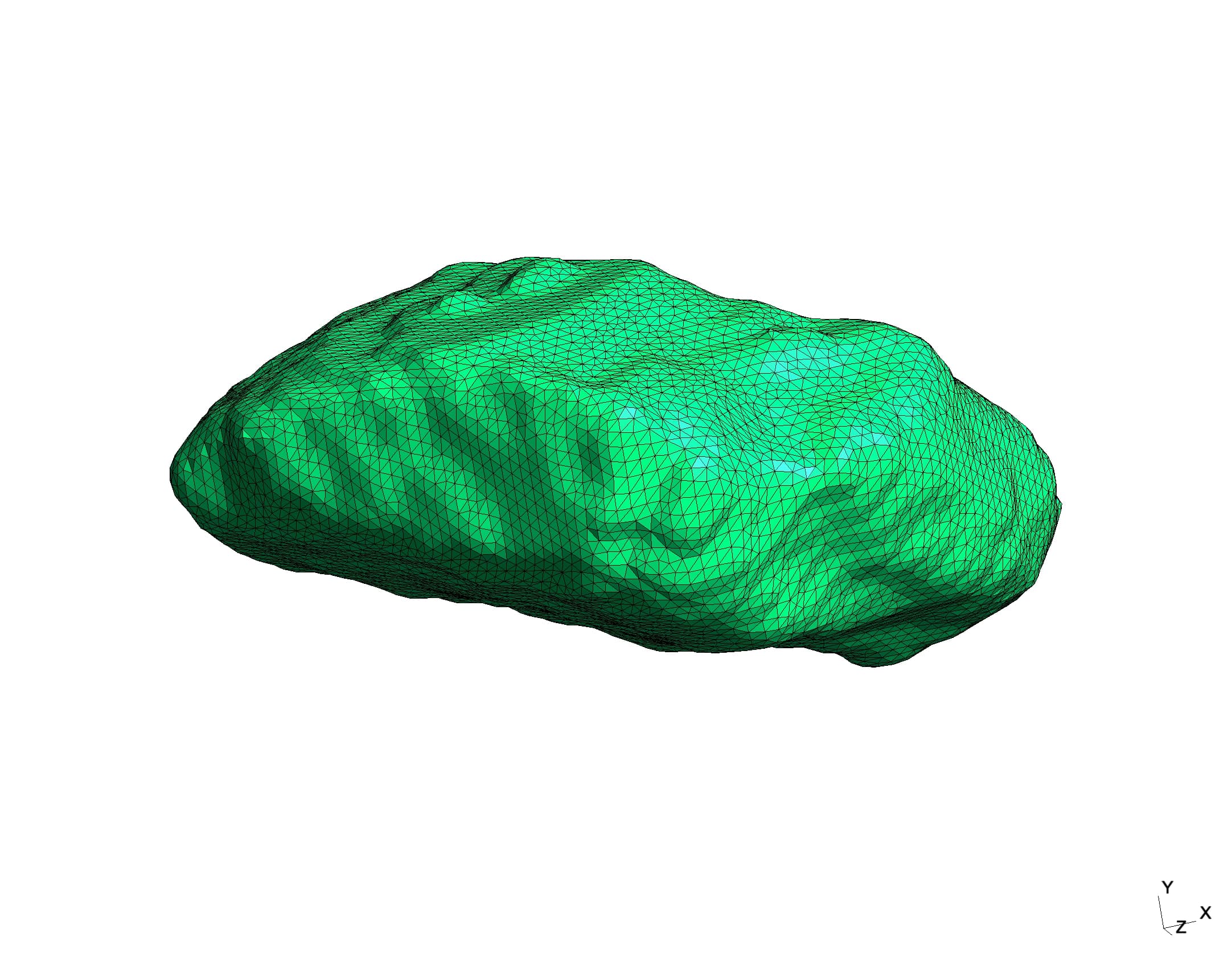}}
        \subfloat[Decimated (factor 20) grain surface mesh]{\includegraphics[width=.32\textwidth]{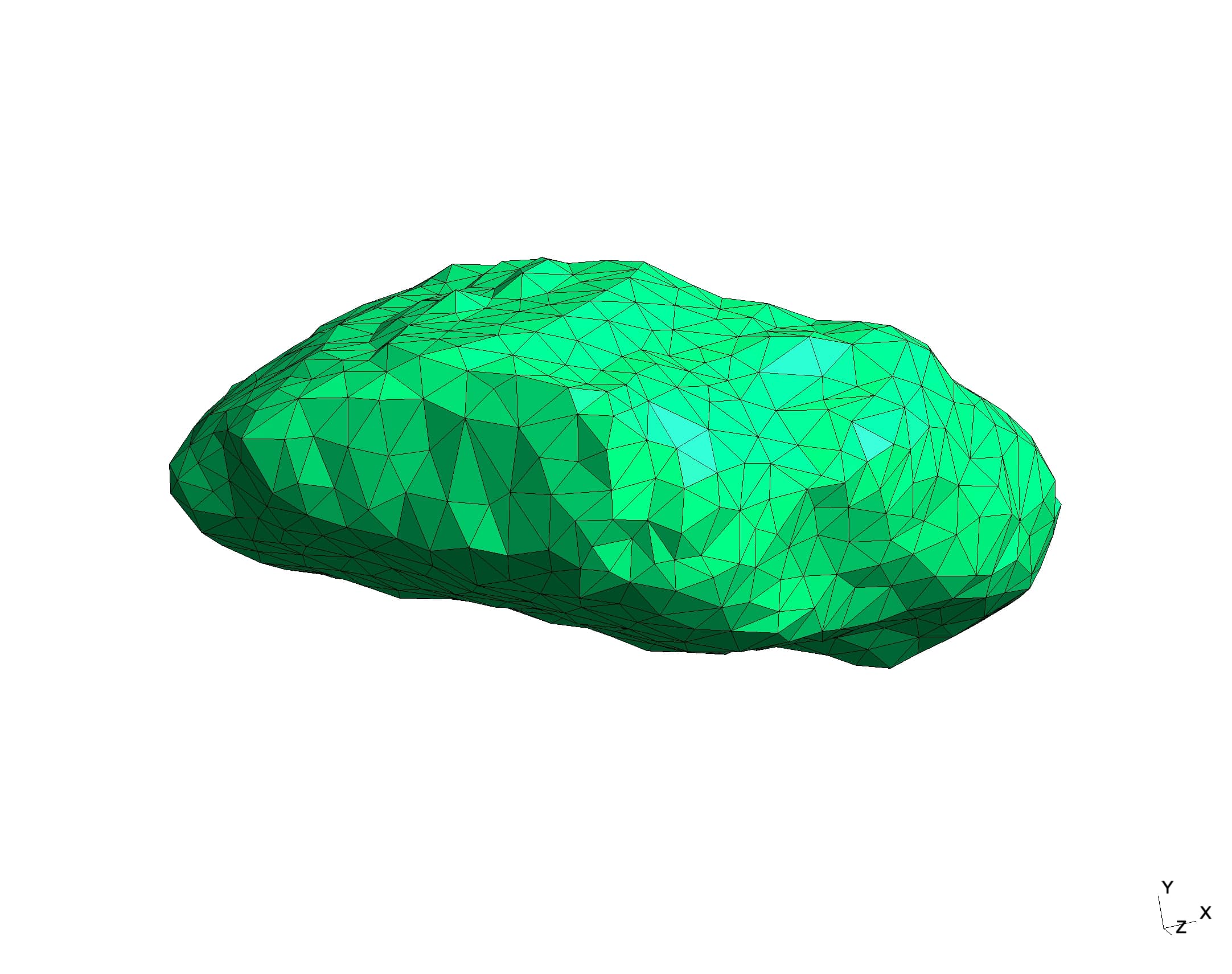}}
        \caption{Post-processing of 3D meshes in Segmentflow. The first panel (a) shows the surface mesh generated from the images using a marching cubes algorithm. The second panel (b) shows the mesh after a smoothing operation. The third panel (c) shows the mesh after it was decimated by a factor of 20.}
        \label{fig:grain-changes}
    \end{figure}

    Figure \ref{fig:grain-changes} illustrates the procedure in which each STL was smoothed and then decimated to reduce the number of resulting tetrahedra in the final mesh. Here, a decimation factor of 20 was used as it resulted in a high-quality mesh yet a spatially converged solution within a computationally tractable number of finite elements. To avoid grain-to-grain contact and possible grain overlap, which were not allowed within the utilized meshing algorithm, STL intersections were detected and removed. This was achieved with $\le 1\%$ loss of IDOX grain volume fraction by isostatically shrinking the affected STLs until separated to a set minimum distance.
    
    In summary, the procedure ensures that a three-dimensional, tetrahedral finite element mesh of a full specimen can be generated with as little loss of IDOX grain volume fraction as possible, though some modifications of each STL was required. Despite missing some of the localized strains that could develop during grain-to-grain contact, the simulations were still able to capture the overall behavior of the experimental data well, since the calibration effort focused on material properties of the matrix taken as a homogeneous mixture of small IDOX grains and Estane (Section \ref{sec:baseline-sim}).

\begin{figure}[!h]
        \centering
        \subfloat[IDOX grains rendered without matrix]{\includegraphics[width=.5\textwidth]{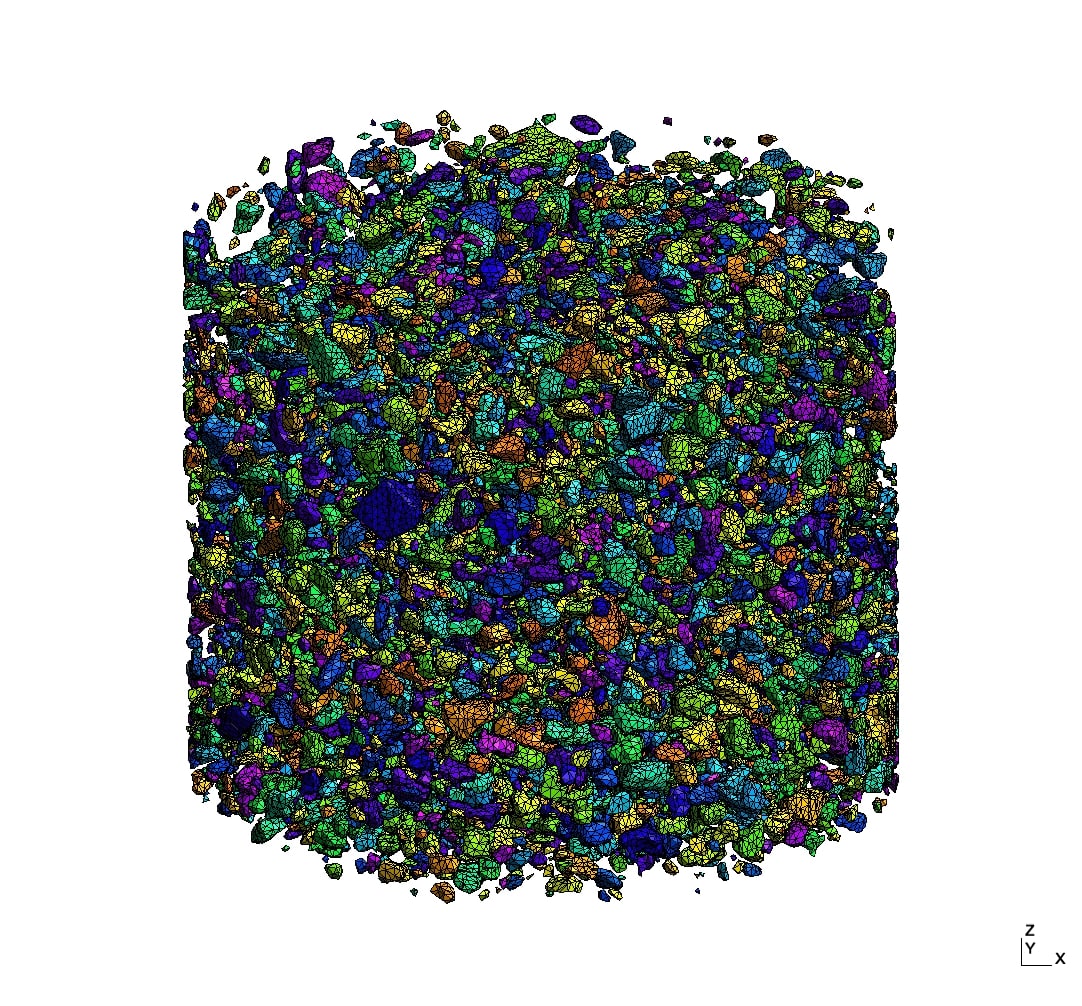}}
        \subfloat[Full tetrahedral mesh]{\includegraphics[width=.5\textwidth]{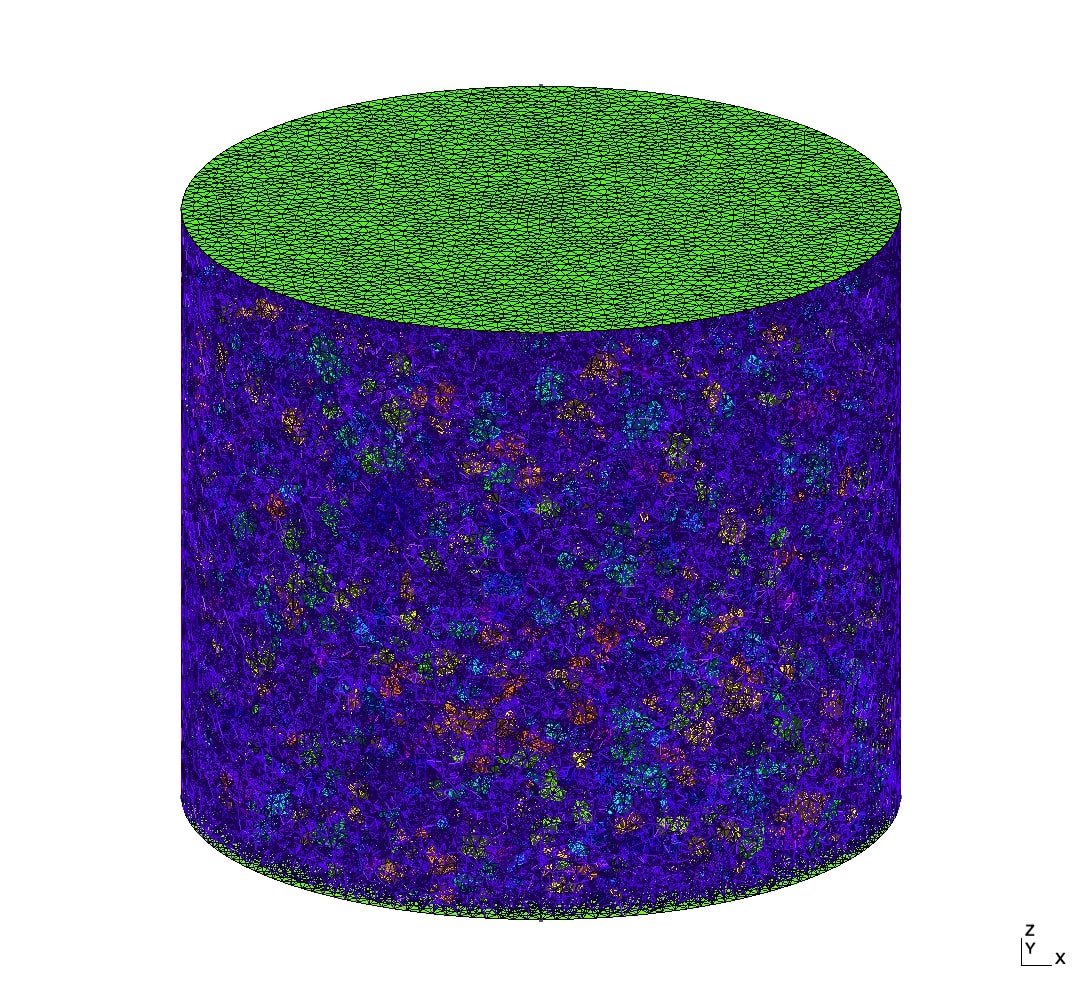}}
        \caption{M-50-01 mesh. The colorful shapes in the interior are the IDOX grains, and the more solid purple is the matrix.}
        \label{fig:meshes}
    \end{figure}

    A cylinder aligned with the loading axis was then fitted around the STL cluster, and a tetrahedral mesh was generated using Gmsh \cite{GeuzaineGmsh2009}. Meshes were created based on the three CT-scanned specimens: S-50-02, S-50-03, and S-50-01. The resulting meshes are labeled M-50-02, M-50-03, and M-50-01, respectively. Details are provided in Table \ref{tab:mesh-values}, and renderings of the M-50-01 mesh are shown in Figure \ref{fig:meshes}. 

    \begin{table}[h!]
    \centering
    \begin{tabular}{lrrr}
        \hline
        & M-50-02 & M-50-03 & M-50-01 \\
        \hline
        \hline
        Sample height ($\si{\milli\meter}$) & 4.67 & 5.12 & 4.95 \\
        Sample radius ($\si{\milli\meter}$) & 2.5 & 2.5 & 2.5 \\
        Sample volume ($\si{\milli\meter\cubed}$) & 91.70 & 100.53 & 97.19 \\
        Mesh height ($\si{\milli\meter}$) & 4.1 & 4.5 & 4.4 \\
        Mesh radius ($\si{\milli\meter}$) & 2.5 & 2.5 & 2.5 \\
        Mesh volume ($\si{\milli\meter\cubed}$) & 82.54 & 91.34 & 88.49 \\
        Voxels of STLs ($\mathrm{x}10^6$) & 8.9 & 11.3 & 13.5 \\ 
        Particles meshed & 8302 & 10104 & 11171 \\
        Grain volume ($\si{\milli\meter\cubed}$) & 15.66 & 19.90 & 23.72 \\
        Grain volume fraction & 0.190 & 0.218 & 0.268 \\
        Elements ($\mathrm{x}10^6$) & 2.30 & 2.69 & 2.86 \\
        Degrees of freedom ($\mathrm{x}10^6$) & 12.3 & 14.2 & 15.1 \\
        \hline
    \end{tabular}
    \caption{Details for the direct numerical simulation geometry and meshes derived from CT scans of three manufactured specimens.}
    \label{tab:mesh-values}
    \end{table}

\subsection{Simulation Setup and Constitutive Model Overview}
\label{sec:FESetup}

   The Ratel solid mechanics library \cite{ratel-user-manual} was used to simulate the quasi-static compression of the specimens. This performance-portable software leverages matrix-free methods and p-multigrid preconditioning to enable efficient and scalable numerical simulations \cite{BrownPerformance2022, ratel-joss-paper}. For both the IDOX grains and the matrix, the finite-strain phase-field model of fracture implemented in Ratel \cite{ratel-user-manual} and described in \ref{sec:app-materialmodel} was used. As in small-strain, isotropic elasticity, the undamaged elastic response depends on only two elastic constants, here Young's modulus $E$ and Poisson's ratio $\nu$. The resistance of the material to damage nucleation and propagation is instead governed by the critical energy release rate $G_c$ and the damage viscosity $\xi$. With the residual stiffness $\eta$ (see \ref{sec:app-materialmodel}), the latter also serves as a regularization parameter that helps prevent numerical instabilities \cite{TanneModeling2018}. To model the action of the lubricated platens, frictional contact boundary conditions, using a Coulomb friction model with viscous damping \cite{MarquesSurvey2016, MlikaUnbiased2017}, were applied to the top and bottom surfaces of the cylindrical domain. This allowed the mesh to realistically bulge radially at the center, given the applied $220\ \si{\micro\meter}$ of axial displacement while preventing unrealistic stress concentrations at the cylinder edges if fully fixed displacement boundary conditions were applied. 
   
\subsection{Baseline Simulation}
\label{sec:baseline-sim}

    A baseline simulation was developed to determine physically reasonable initial estimates for the parameters and confirm that the experimental data could be replicated with the DNS setup. To specify the coupled elastic-damage model, values for the Young's modulus $E$, Poisson ratio $\nu$, critical release rate $G_c$, damage viscosity $\xi$, and residual stiffness $\eta$ were needed for the IDOX grains and matrix. The values for these used in the baseline simulation are given in Table \ref{tab:baseline-values} and were determined through a combination of previous experimental work on IDOX \cite{BurchNanoindentation2017} and test simulations. 
    
    \begin{table}[h!]
    \centering
    \begin{tabular}{llrll}
        \hline
        Parameter Name & Variable & Value & Unit & Calibrated \\ 
        \hline
        \hline
        IDOX Young's modulus & $E^{(I)}$ & 22000 & $\si{\mega\pascal}$ & - \\
        IDOX Poisson ratio & $\nu^{(I)}$ & 0.25 & - & - \\
        IDOX critical release rate & $G^{(I)}_c$ & 1830 & $\si{\pico\joule\per\micro\meter\squared}$ & - \\
        IDOX damage viscosity & $\xi^{(I)}$ & 0.1 & $\si[inter-unit-product = \ensuremath{{}\cdot{}}]{\mega\pascal\second}$ & - \\
        IDOX residual stiffness factor & $\eta^{(I)}$ & 0.01 & - & - \\
        Matrix Young's modulus & $E^{(M)}$ & 275 & $\si{\mega\pascal}$ & * \\
        Matrix Poisson ratio & $\nu^{(M)}$ & 0.25 & - & - \\
        Matrix critical release rate & $G^{(M)}_c$ & 2.5 & $\si{\pico\joule\per\micro\meter\squared}$ & * \\
        Matrix damage viscosity & $\xi^{(M)}$ & 0.3 & $\si[inter-unit-product = \ensuremath{{}\cdot{}}]{\mega\pascal\second}$ & * \\
        Matrix residual stiffness factor & $\eta^{(M)}$ & 0.01 & - & - \\
        Coulomb model friction coefficient & $\mu$ & 0 & - & - \\
        Coulomb model viscous damping & $F_v$ & 0.01 & $\si[inter-unit-product = \ensuremath{{}\cdot{}}]{\newton\meter\per\second}$ & - \\ 
        Phase-field model length scale & $l_0$ & 0.06 & $\si{\milli\meter}$ & - \\
        \hline
    \end{tabular}
    \caption{Baseline simulation parameters. In the last column, the parameters that are calibrated are indicated with * and the other parameters are held constant.}
    \label{tab:baseline-values}
    \end{table}

\begin{figure}[h!]
        \centering
        \subfloat[Baseline simulation force-displacement data (black) and the $50\ \si{\degreeCelsius}$ experimental data (colors)]{\includegraphics[width=.75\textwidth]{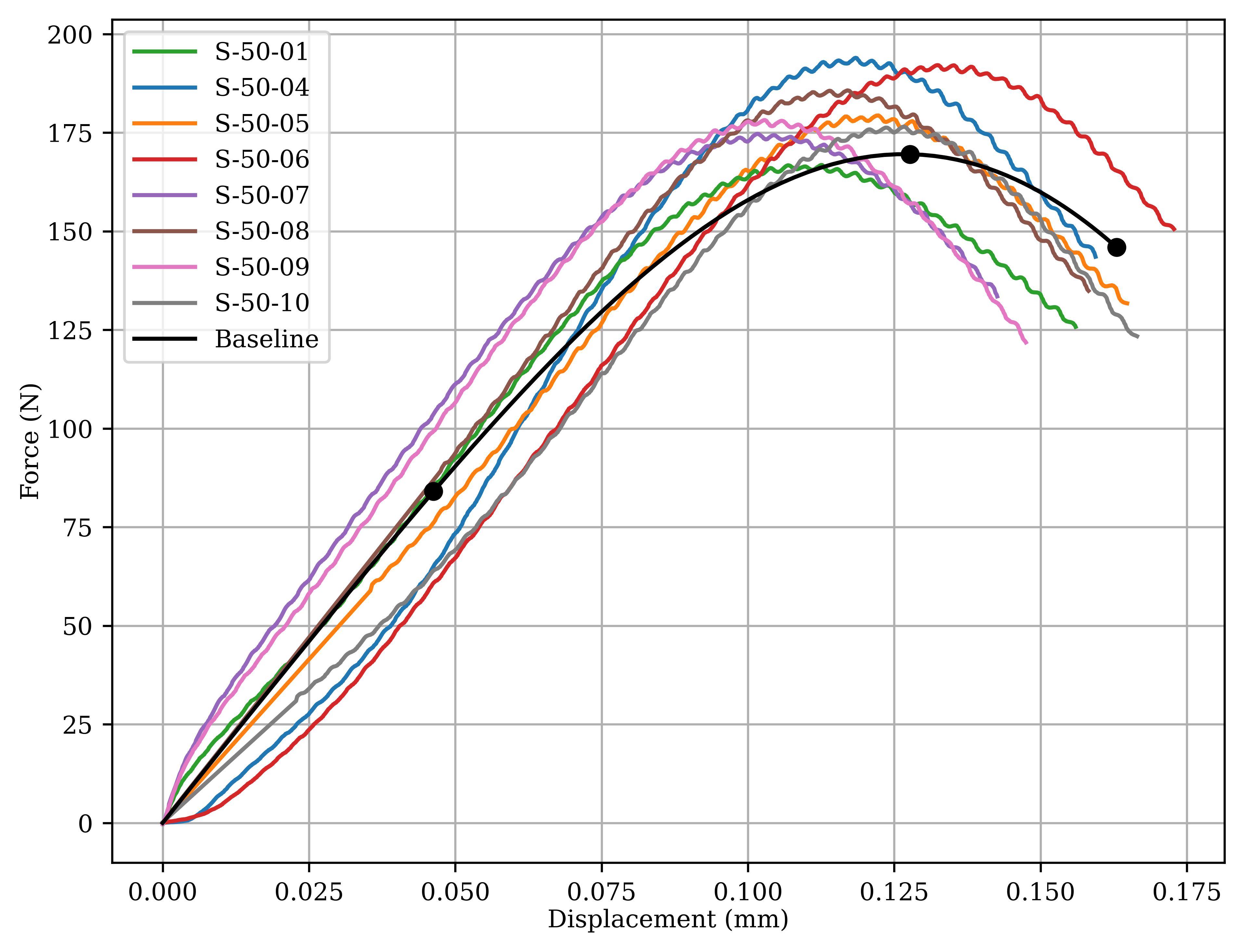}} \\
        \subfloat[Damage field at $0.046\ \si{\milli\meter}$]{\includegraphics[width=.32\textwidth]{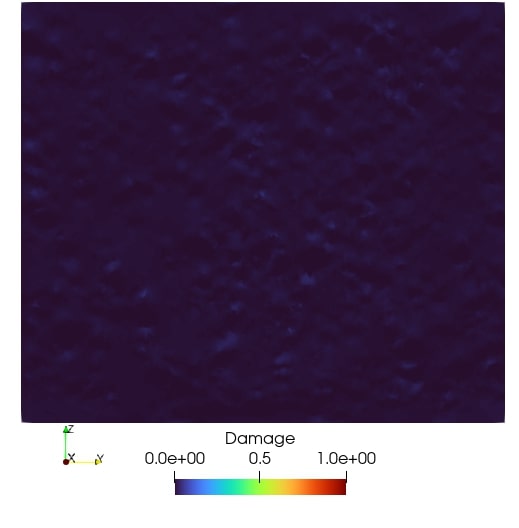}}
        \subfloat[Damage field at $0.123\ \si{\milli\meter}$]{\includegraphics[width=.32\textwidth]{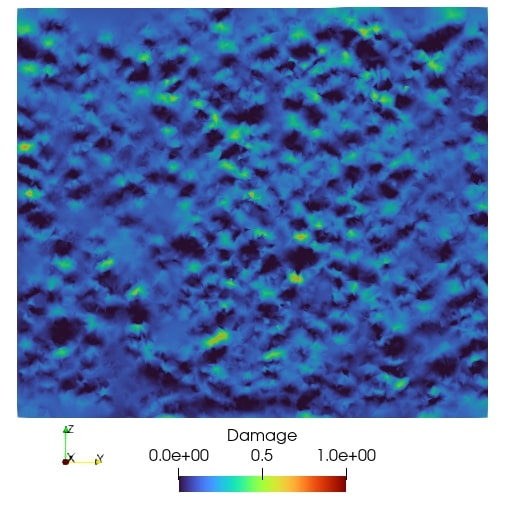}}
        \subfloat[Damage field at $0.163\ \si{\milli\meter}$]{\includegraphics[width=.32\textwidth]{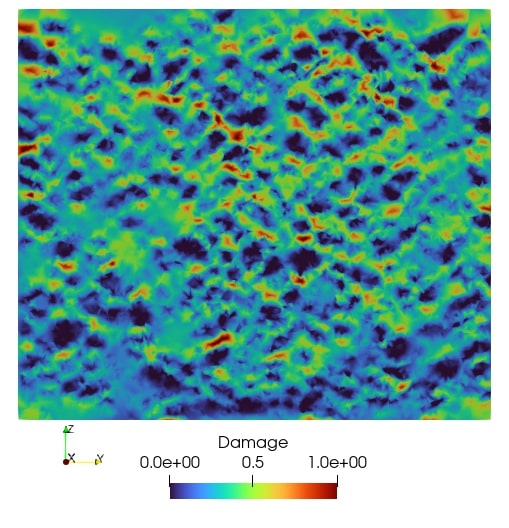}}
        \caption{Ratel-FEM baseline DNS force-displacement curve and damage field at three displacements with the M-50-01 mesh.}
        \label{fig:baseline}
    \end{figure}

    The global force-displacement results of the baseline simulation and experimentally obtained force-displacement curves are plotted in Figure \ref{fig:baseline} (a). In general, all of the curves produced similar slopes in the initial, elastic response. The slope of the curves gradually decreased as damage accumulated, and the peak force was approached at the onset of cracking. At this stage in the applied deformation, damage nucleation sites appeared within the matrix and near IDOX grains. Specifically, they aligned with the loading axis and were located near grain boundaries that created stress concentration. Consistent with experimental observations, the force decreased gradually post-peak. The absence of sharp load drops can be attributed to IDOX grains hindering long-range crack propagation. Renderings of the damage contours that developed during the simulation as seen through the middle of the simulation can be found in Figure \ref{fig:baseline}; the first rendering (b) was taken roughly midway through the elastic response, the second rendering (c) was at peak-force, and the final rendering (d) shows the post-peak accumulation of damage. Qualitatively, these echo the damage patterns seen in the experiments as seen in Figure \ref{fig:exp-snapshots}, and it is apparent that localized damage is occurring around the large IDOX grains highlighting the effect of the underlying microstructure on the specimen's behavior. 

\section{Workflow and Computing}
\label{sec:workflow}

    The baseline simulation run with the parameters in Table \ref{tab:baseline-values} was consistent with the experimental data. Though, with more rigorously calibrated values, the simulations could predict the experiments better. To this end, the goal was to calibrate the matrix Young's modulus $E^{(M)}$, critical release rate $G^{(M)}_c$, and damage viscosity $\xi^{(M)}$ using the experimental data. In this work, the IDOX parameters were held constant, as it was found during testing that changing the properties of IDOX did not significantly alter the resulting force-displacement curve. This was expected considering the microstructure of the simulations with the matrix in between each large IDOX grain. The friction boundary condition parameters were also held constant to focus the calibration on the material properties of the matrix.  

    As can be seen in the force-displacement data (Figures \ref{fig:all-exp-data}, \ref{fig:baseline} (a)), the experiments varied in their measured responses. To account for this variation in the data, the calibration was conducted with quantified uncertainty in the form of Bayesian inference. Additionally, by quantifying the uncertainty in the calibrated parameters through the posterior distribution, the resulting distributions over the parameters can be sampled to generate ensembles of calibrated values for downstream modeling efforts. The overall calibration workflow is shown in Figure \ref{fig:workflow}. 

    To perform the Bayesian inference, hundreds of Ratel direct numerical simulations (DNS) were run to populate a database of simulation results for a given set of input material properties. The DNS contained millions of degrees of freedom (see Table \ref{tab:mesh-values}), necessitating the use of high-performance computing (HPC) resources. Each DNS took approximately 2-4 hours to reach post-peak on Lawrence Livermore National Laboratory's Lassen supercomputer using 40 NVIDIA Volta V100 GPUs (12 threads per GPU). 


\section{Calibration with Uncertainty Quantification}
\label{sec:uq}

\subsection{Uncertainty Quantification Framework}

    A Bayesian inference framework was used to calibrate the parameters with uncertainty quantification, where the aim was to find values for the parameters that result in the ``best'' match between the simulated and experimental data according to Bayes' Theorem, 
    \begin{equation}
        P(\boldsymbol{\theta}|\boldsymbol{Y}) \propto P(\boldsymbol{Y}|\boldsymbol{\theta}) P(\boldsymbol{\theta}).
    \end{equation}
    Here, $\boldsymbol{\theta}=[E^{(M)}, G_c^{(M)}, \xi^{(M)}]$ are the parameters to calibrate, $\boldsymbol{Y}$ are the exprimental data (see Section \ref{sec:qois}), $P(\boldsymbol{\theta})$ is the prior distribution over the parameters containing any previous knowledge about them, $P(\boldsymbol{Y}|\boldsymbol{\theta})$ is the likelihood of the data given choices for the parameters, and $P(\boldsymbol{\theta}|\boldsymbol{Y})$ is the posterior distribution which provides the updated information about the parameters after considering the data.

\subsection{Prior Distributions}
\label{sec:priors}

    In this work, the priors were chosen based on the baseline simulation (Section \ref{sec:baseline-sim}) and scoping studies to ensure the simulation results bounded the experimental force-displacement data. Uniform priors were used for all parameters and are listed in Table \ref{tab:priors}.

    \begin{table}[h!]
    \centering
    \begin{tabular}{lllll}
        \hline
        & $E^{(M)}$ ($\si{\mega\pascal}$) & $G_c^{(M)}$ ($\si{\pico\joule\per\micro\meter\squared}$) & $\xi^{(M)}$ ($\si[inter-unit-product = \ensuremath{{}\cdot{}}]{\mega\pascal\second}$) & $N$ samples\\
        \hline
        \hline
        M-50-02 & $U[250,300]$ & $U[0.01, 5.0]$ & $U[0.05, 0.55]$ & 100 \\ 
        M-50-03 & $U[250,350]$ & $U[0.01, 5.0]$ & $U[0.05, 0.55]$ & 200 \\ 
        M-50-01 & $U[250,300]$ & $U[0.01, 5.0]$ & $U[0.05, 0.55]$ & 100 \\ 
        M-50-01-Cal90 & $U[250,450]$ & $U[0.01, 5.0]$ & $U[0.05, 0.85]$ & 100 \\ 
        \hline
    \end{tabular}
    \caption{Uniform priors. }
    \label{tab:priors}
    \end{table}
    
\begin{figure}[!h]
        \centering
        \subfloat[M-50-01 and M-50-02]{\includegraphics[scale=1.0]{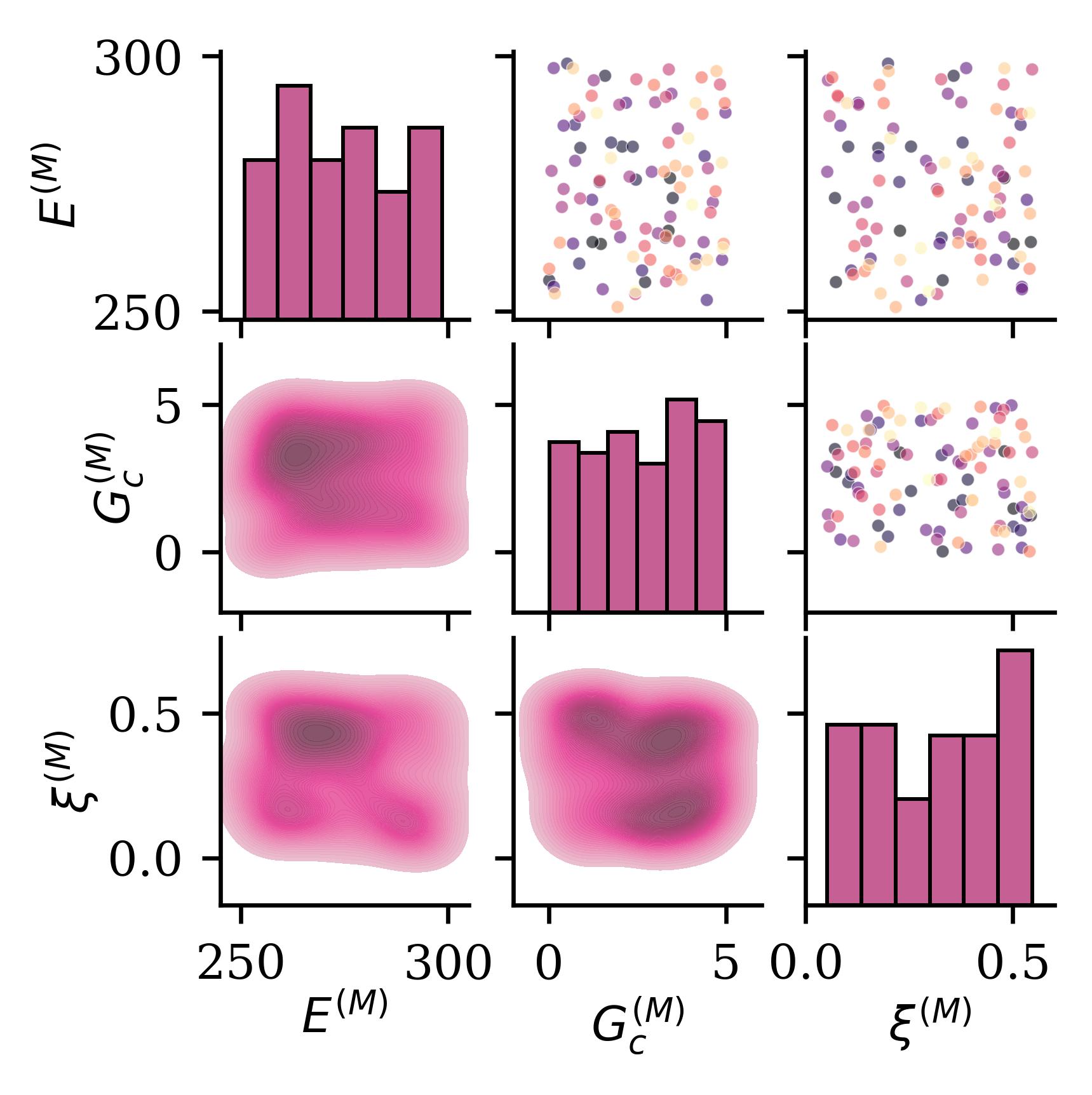}}
        \subfloat[M-50-03]{\includegraphics[scale=1.0]{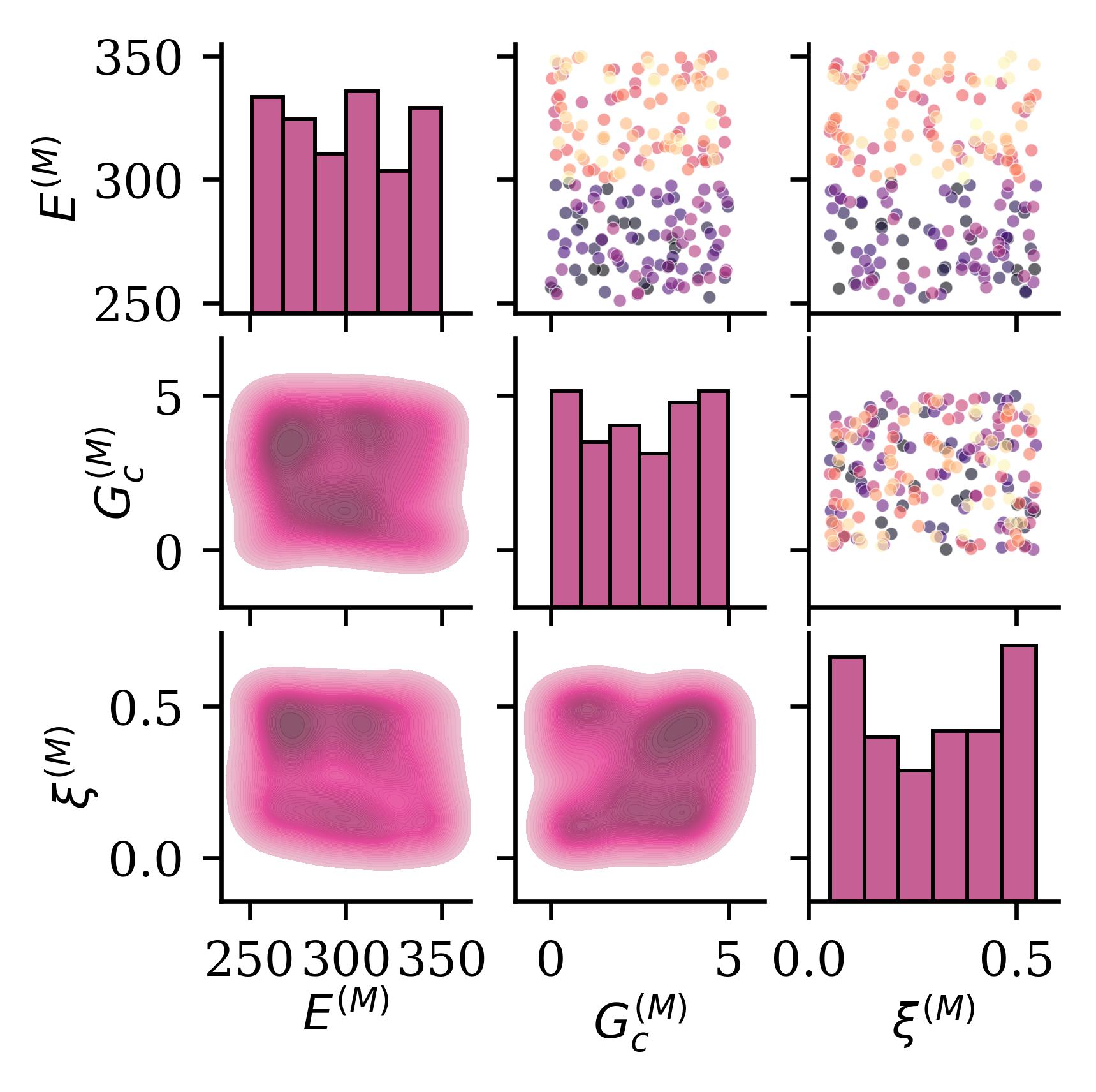}} \\
        \subfloat[M-50-01-Cal90]{\includegraphics[scale=1.0]{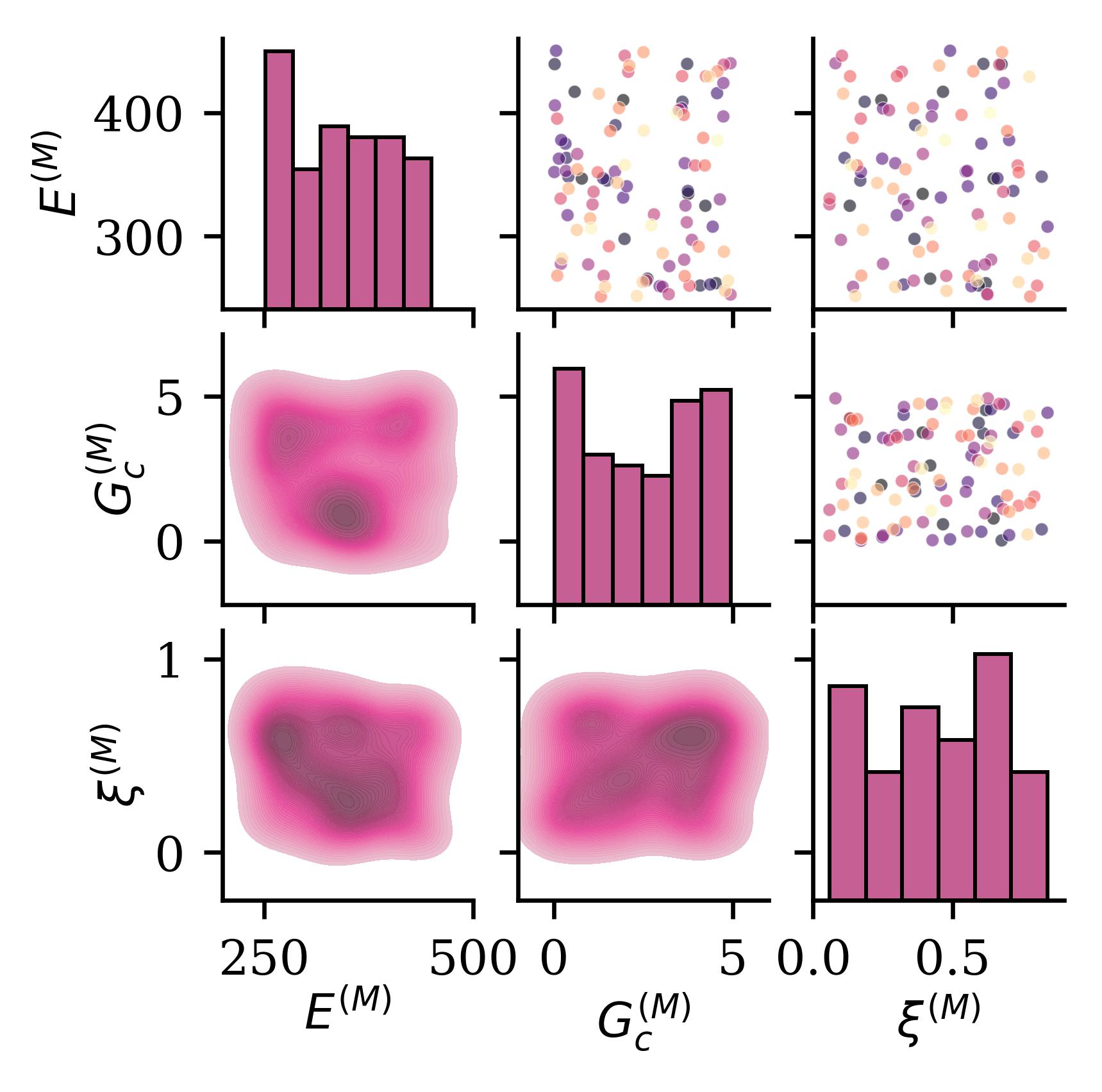}}
        \caption{Prior distribution samples. The histograms on the diagonal show the marginal distribution for each parameter. The upper triangle shows pairwise scatterplots, and the lower triangle shows the corresponding kernel density estimates. }
        \label{fig:priors}
    \end{figure}

    Each prior was independently sampled 100 times to form sets of parameters with which to run the Ratel simulations (as described in Section \ref{sec:workflow}). These samples are shown in Figure \ref{fig:priors}. Note that for the M-50-03 mesh, with the same prior as M-50-01 and M-50-02, the resulting marginal posterior for $E^{(M)}$ was truncated by the upper bound of the distribution. Therefore, for M-50-03, the prior on $E^{(M)}$ was extended to $U[250,350]$ by taking an additional 100 samples from the prior $E^{(M)} \sim U[300,350] \times G^{(M)}_c \sim U[0.01, 5.0] \times \xi^{(M)} \sim U[0.05, 0.55]$ and considering both the additional 100 samples and the original 100 samples. 

\subsection{Quantities of Interest}
\label{sec:qois}

\begin{figure}[!h]
        \centering
        \subfloat[$50\ \si{\degreeCelsius}$]{\includegraphics[width=0.49\linewidth]{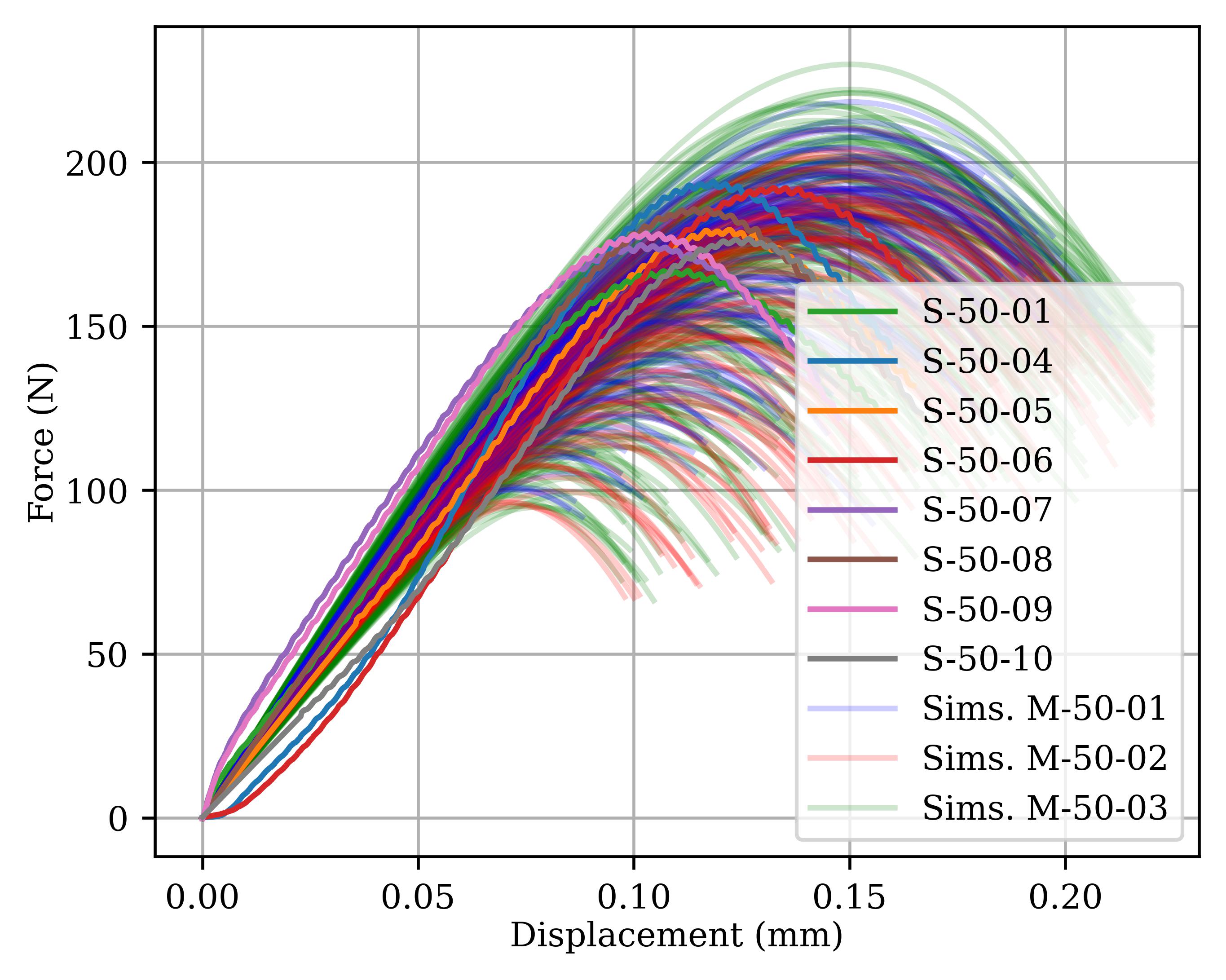}}
        \hfill
        \subfloat[$90\ \si{\degreeCelsius}$]{\includegraphics[width=0.49\linewidth]{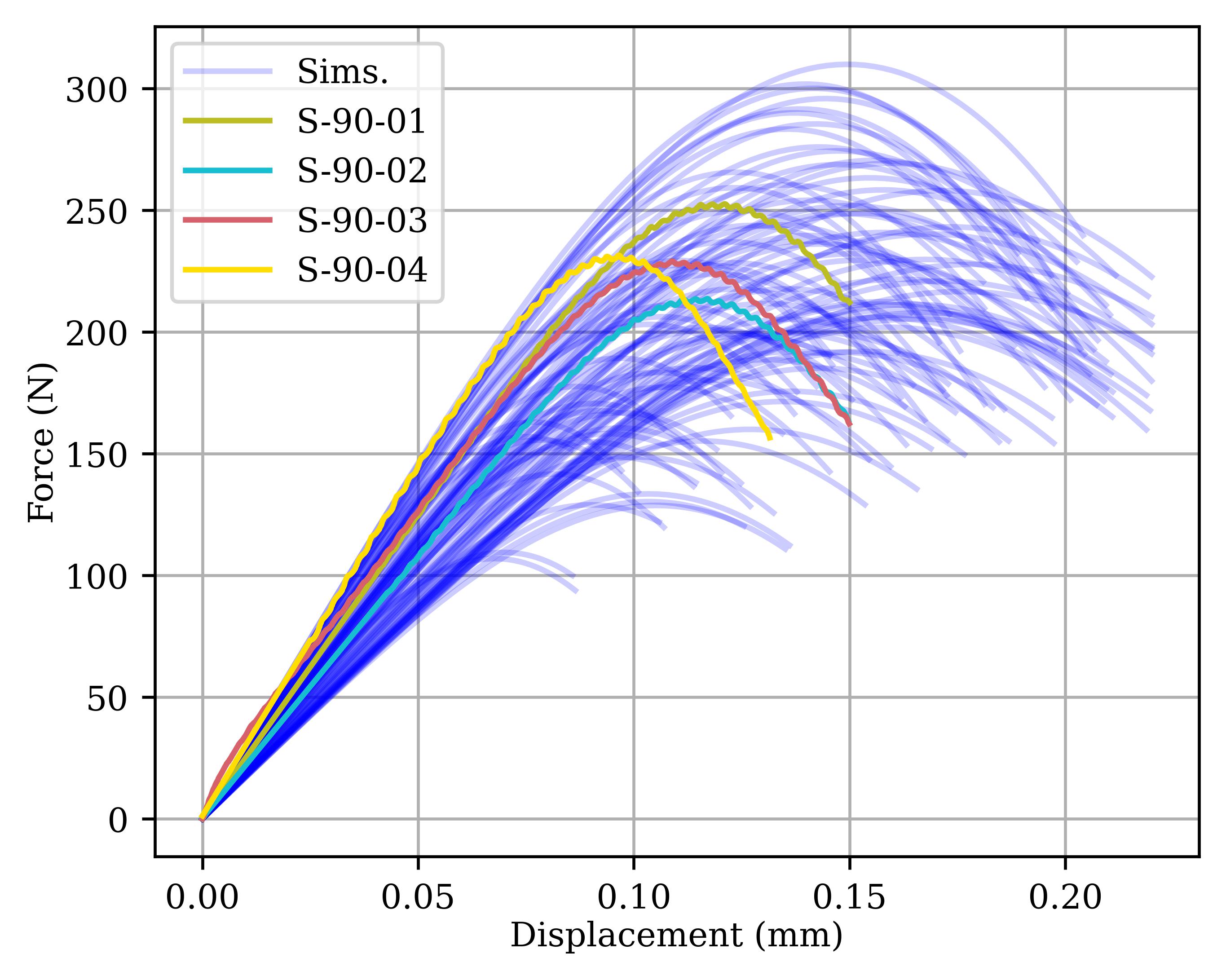}}  
        \caption{Experimental and simulated force-displacement data.}
        \label{fig:force-dispalacement-all}
    \end{figure}

    With the samples from the prior, depending on the case, 100 or 200 DNS were run to generate simulated force-displacement curves. These data are shown with the experimental force-displacement curves in Figure \ref{fig:force-dispalacement-all}. The simulated force-displacement curves bound the experimental values, indicating that the choices of priors were broad enough to capture the experimental variation. 

\begin{figure}[!h]
        \centering
        \subfloat[$50\ \si{\degreeCelsius}$]{\includegraphics[width=0.98\linewidth]{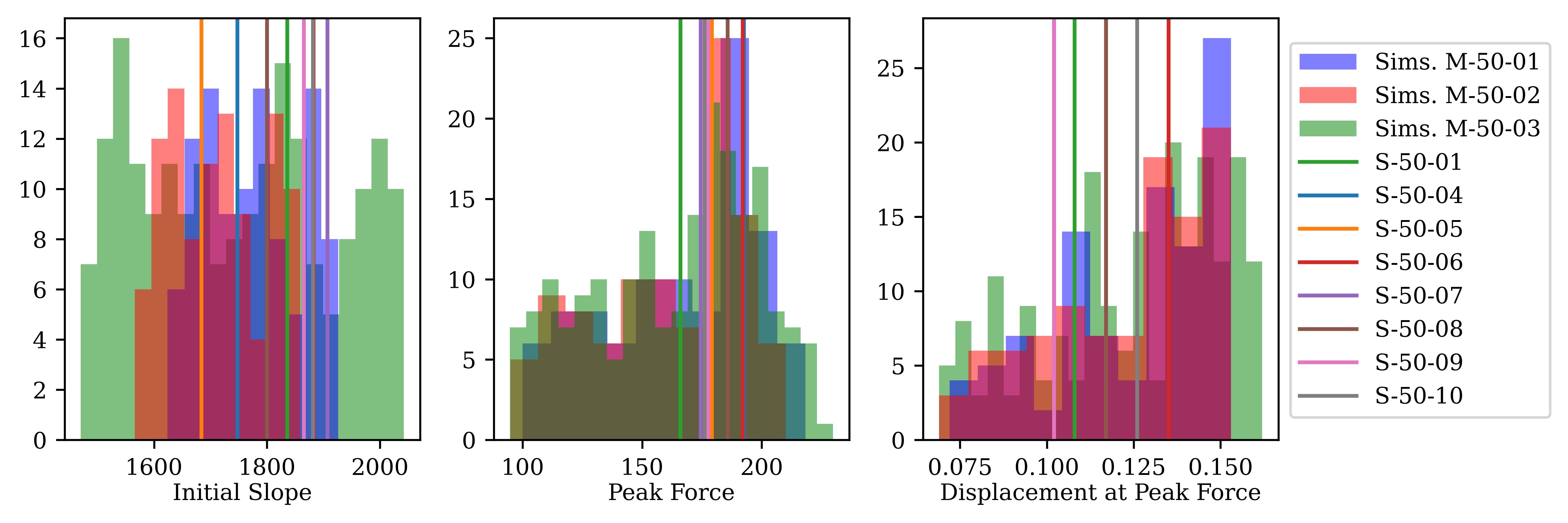}}\\
        \subfloat[$90\ \si{\degreeCelsius}$]{\includegraphics[width=0.98\linewidth]{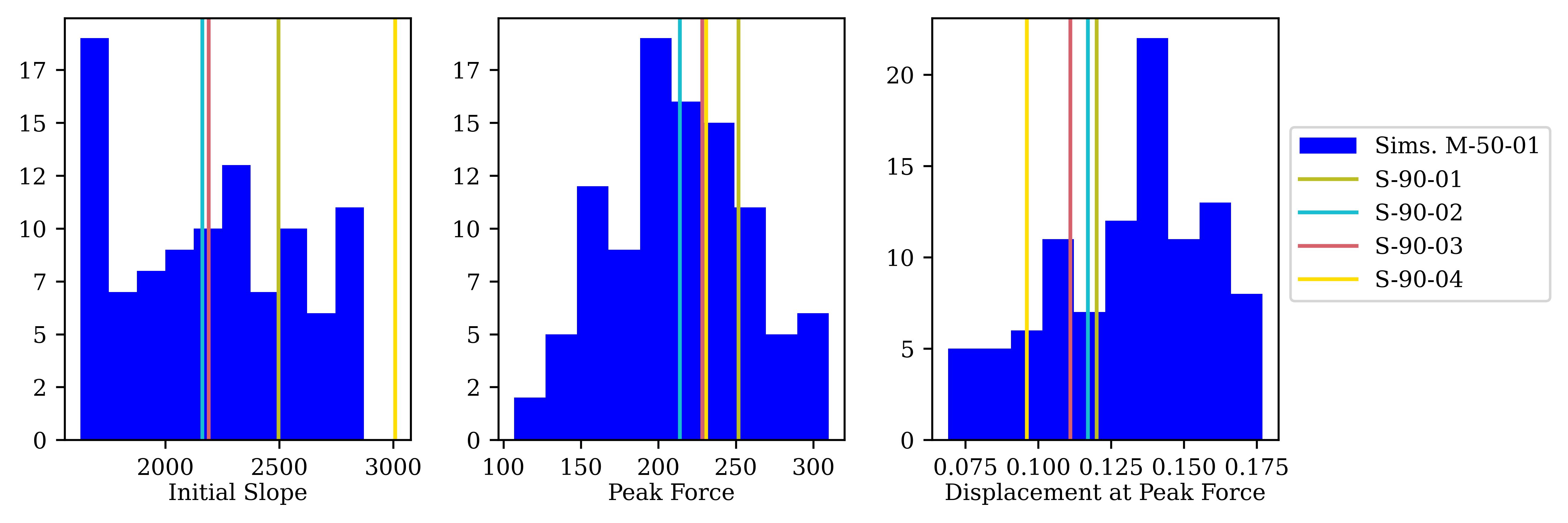}}        
        \caption{Experimental and simulated QoIs computed from the force-displacement data shown in Figure \ref{fig:force-dispalacement-all}. }
        \label{fig:qois-all}
    \end{figure}

    Instead of using the entire force-displacement curve as the data $\boldsymbol{Y}$, several quantities of interest (QoIs) were computed from the experimental and simulated force-displacement curves. These QoIs were the initial slope $m$, peak force $F$, and displacement at peak force $d_F$. The peak force and displacement at peak force are the value and location of the maximum value for each of the force-displacement curves. The initial slope was computed as the slope of the curve where the behavior was most linear. For the simulations, S-50-04, and four S-90- experimental curves, this was as $0.3d_F$. For the other seven $50\ \si{\degreeCelsius}$ curves, it was at $0.55d_F$. Figure \ref{fig:qois-all} shows the QoIs from the simulation data compared to the experimental QoIs. Overall, the simulated and experimental QoIs overlap or are close for all quantities; therefore, are appropriate to use for calibration.

\subsection{Likelihood and Surrogate Model}
\label{sec:likelihood_pce}

    The assumption is made that the experiments are independent, and, therefore, the sets of QoIs from each experiment are independent. This means that the likelihood can be evaluated as the product of the likelihoods for each experimental data set, 
    \begin{equation}
        P(\boldsymbol{Y}|\boldsymbol{\theta}) = \prod_{k=1}^{n_{exp}} P(\boldsymbol{y}_k | \boldsymbol{\theta}),
    \end{equation}
    where $\boldsymbol{y}_k = [m, F, d_F]\in \mathbb{R}^{n_k}$ is the vector of the QoIs computed from each experimental force-displacement curve.

    For the likelihood of each experimental data set, the discrepancy between the experimental and simulated QoIs was modeled as coming from a multivariate normal distribution 
    \begin{equation}
        P(\boldsymbol{y}_k | \boldsymbol{\theta}) = (2\pi)^{-n_k /2} \vert \boldsymbol{\Sigma}\vert^{-1/2} \exp \bigg( -\frac{1}{2} \left(\boldsymbol{y}_k - \boldsymbol{m}(\boldsymbol{\theta})\right)^T \boldsymbol{\Sigma}^{-1} \left(\boldsymbol{y}_k - \boldsymbol{m}(\boldsymbol{\theta})\right) \bigg),
    \end{equation}
    where $n_k=3$ is the number of QoIs, $\boldsymbol{m}(\boldsymbol{\theta})$ is the vector of QoIs computed from a Ratel simulation run with input parameters $\boldsymbol{\theta}$, and $\boldsymbol{\Sigma}$ is the covariance matrix. 

    In practice it can be difficult to compute the inverse of $\boldsymbol{\Sigma}$, so here the Cholesky decomposition of the covariance matrix $\boldsymbol{\Sigma} = \boldsymbol{LL}^T$ was used instead, where $\boldsymbol{L}$ is a lower triangular matrix \cite{GelmanBayesian2013}. The likelihood can then be written as, 
    \begin{equation}
        P(\boldsymbol{y}_k | \boldsymbol{\theta}) = (2\pi)^{-n_k /2} \vert \boldsymbol{LL}^T\vert^{-1/2} \exp \bigg( -\frac{1}{2} \left(\boldsymbol{y}_k - \boldsymbol{m}(\boldsymbol{\theta})\right)^T (\boldsymbol{LL}^T)^{-1} \left(\boldsymbol{y}_k - \boldsymbol{m}(\boldsymbol{\theta})\right) \bigg).
    \end{equation}
    The Cholesky decomposed covariance matrices were modeled such that the correlation matrices were drawn from an Lewandowski-Kurowicka-Joe (LKJ) distribution \cite{GelmanBayesian2013} with $\eta= 3.0$ and standard deviations following a $\mathcal{N}(0.15, 0.25)$ distribution, using the functions in the Python package PyMC \cite{PyMC2023}. 
    
    As discussed in Section \ref{sec:workflow}, each Ratel DNS took about 2-4 hours to run on 40 NVIDIA Volta GPUs, making it a computationally prohibitive model to use in evaluating sampled parameter sets. A common approach to accelerate Bayesian inference with expensive models is to employ a surrogate model which maps values for the parameters to the QoIs using a model form that is faster to evaluate than the original \cite{MarzoukStochastic2007, LiAdaptive2014, WangUsing2005, ChristenMarkov2005, HeidenreichBayesian2015, AlsupContext2023}. Here, the surrogate model is in the form of a polynomial chaos expansion (PCE) \cite{WienerHomogeneous1938, XiuWeiner2002, XiuModeling2003, GhanemStochastic2003, DoostanNon2011, HamptonCompressive2015}, which does the mapping according to,
    \begin{equation}
        \boldsymbol{m}(\boldsymbol{\theta}) = \sum_{j=1}^\infty \boldsymbol{c}_j \psi_j(\boldsymbol{\theta}),
    \end{equation}
    where $\boldsymbol{\theta}$ is the d-dimensional vector of random inputs ($d=3$) with joint probability density function $f(\boldsymbol{\theta})$ and set of possible realizations $\Omega$, $\boldsymbol{m} = [m, F, d_F] \in \mathbb{R}^{M}$ ($M=3$) is the QoI vector, $\psi_j(\boldsymbol{\theta})$ is a multivariate orthogonal polynomial evaluated at the random inputs $\boldsymbol{\theta}$, and the vector $\boldsymbol{c}_j \in \mathbb{R}^M$ contains deterministic coefficients. Note, the entries of $\boldsymbol{\theta}$ are independent and identically distributed (i.i.d) and were normalized to be $[-1, 1]$. Since the input parameters were sampled from uniform distributions, multivariate Legendre polynomials were used for $\psi_j$ \cite{XiuWeiner2002}. 

    In practice, the expansion is truncated to 
    \begin{equation}
        \boldsymbol{m}(\boldsymbol{\theta}) \approx \sum_{j=1}^P \boldsymbol{c}_j \psi_j(\boldsymbol{\theta}),
    \end{equation}
    with total order $p$, where $\boldsymbol{j} = (j_1, \cdots, j_d)$ such that $\Vert\boldsymbol{j}\Vert_1 \le p$. Such an expansion has $P= (p+d)!/(p!d!)$ basis functions. 
    
    In this work, to determine the total order $p$ to use, the PCE was computed for orders 0 to 8 using a training set of 70\% of the simulation data. A training error vs.~validation error split was then used to identify the lowest order of PCE that generated accurate predictions of the QoIs while not overfitting. This order of PCE was then re-fit using all the simulation data and used as the surrogate model to the Ratel simulation QoIs in the computation of the likelihood. A PCE of order three ($p=3$) was used for all cases considered.

    The coefficients are collected into the matrix $\boldsymbol{C} = [\boldsymbol{c}_1, \cdots, \boldsymbol{c}_P] \in \mathbb{R}^{M\times P}$ and to determine $\boldsymbol{C}$, a regression problem with samples of the QoI was constructed. Let individual realizations of $\boldsymbol{\theta}$ be denoted as $\boldsymbol{\theta}^{(i)}$ and consider a set of $N$ input samples as $\{\boldsymbol{\theta}^{(i)} \}^N_{i=1}$ and corresponding QoI samples $\{\mathcal{\boldsymbol{m}}(\boldsymbol{\theta}^{(i)})  \}^N_{i=1}$ organized in a data matrix $\boldsymbol{{M}} = [\mathcal{\boldsymbol{m}}(\boldsymbol{\theta}^{(1)}), \cdots, \mathcal{\boldsymbol{m}}(\boldsymbol{\theta}^{(N)})] \in \mathbb{R}^{M \times N}$. Then, the goal is to solve for $\boldsymbol{C}$ in the linear system
    \begin{equation}
        \boldsymbol{C\Psi} \approx \boldsymbol{M},      
    \end{equation}
    where $\boldsymbol{\Psi}(j,i) = \psi_j(\boldsymbol{\theta}^{(i)})$. Here, $N=100$ or $200$ and $P=20$, so the regression problem is over determined with $N > P$. Thus, least squares was used to solve for the coefficient matrix $\boldsymbol{C}$ \cite{HadigolLeast2018, HamptonCoherence2015, HosderNon2006} with $N$ sufficient to compute $\boldsymbol{C}$ \cite{HamptonCompressive2015, HadigolLeast2018}.  
    
\subsection{Computing the Posterior Distributions}
\label{sec:mcmc}
    
    The Bayesian framework used here was implemented with the Python package PyMC \cite{PyMC2023} and the posterior distribution was sampled with Markov Chain Monte Carlo (MCMC). The MCMC sampling with PyMC used 4 chains, 50,000 samples per chain, and an initial tuning period of 2,000 steps. The chains were post-processed with a burn-in period of 25,000 steps, and every tenth sample in the chains was used to reduce the correlation between steps. The final posterior distribution for each calibration is reported in Section \ref{sec:results} with 10,000 samples and summarized via the maximum {\it a posteriori} (MAP) value. 


\section{Results and Discussion}
\label{sec:results}

    \subsection{Calibration with $50\ \si{\degreeCelsius}$ Data}
    \label{sec:results-50C}

\begin{figure}[!h]
        \centering
        \includegraphics[scale=1.1]{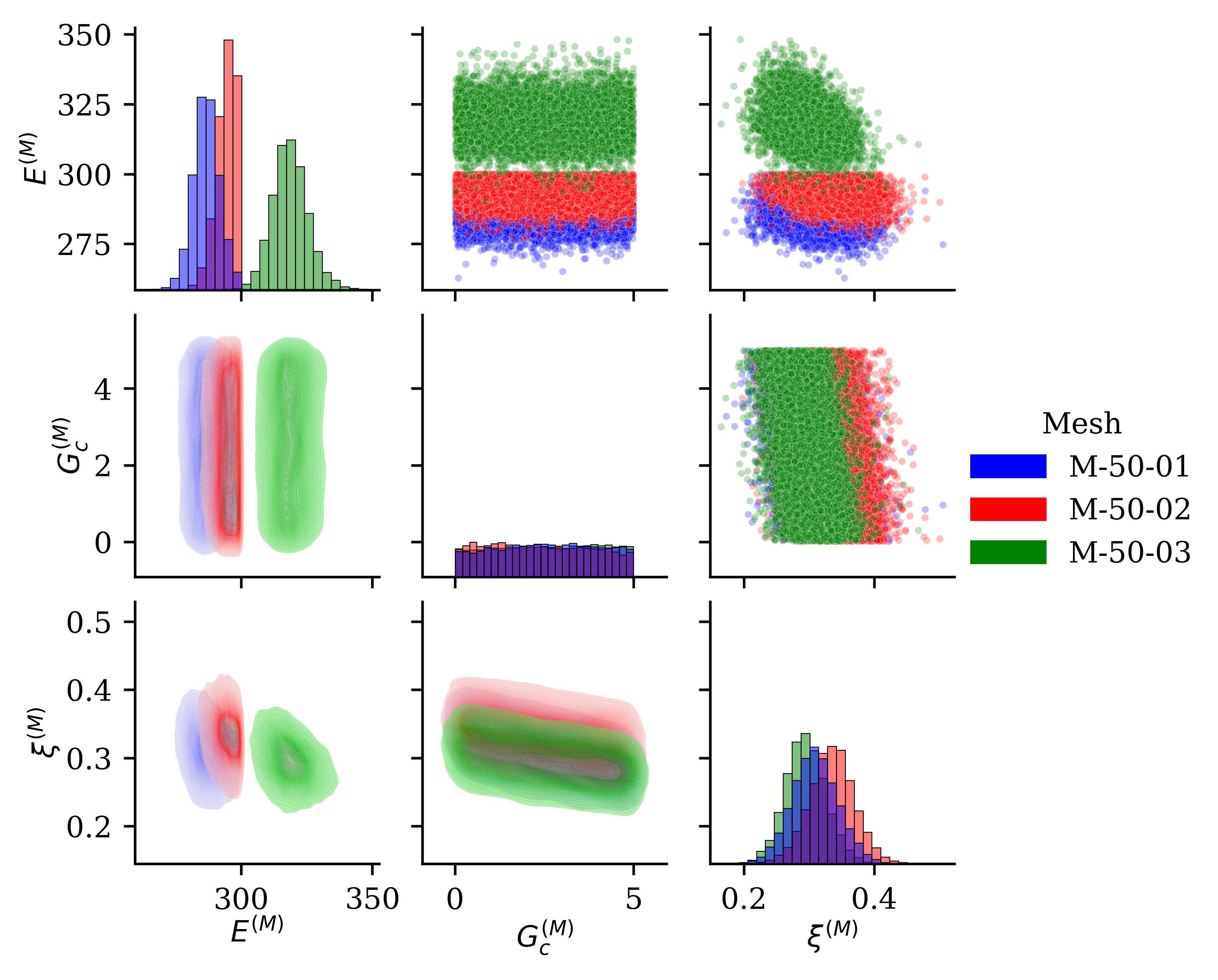}
        \caption{Posteriors for $50\ \si{\degreeCelsius}$ calibrations}
        \label{fig:posteriors-50C}
    \end{figure}

    The calibration framework described in Section \ref{sec:uq} was applied to the $50\ \si{\degreeCelsius}$ experimental data, considering the three meshes derived from CT data. The resulting posterior distributions are shown in Figure \ref{fig:posteriors-50C} and the MAP values are reported in Table \ref{tab:calibration-MAPvalues}. 

    All three initial conditions follow the same pattern from the prior to the posterior in which the marginal distributions over Young's modulus and damage viscosity move from uniform to having a defined peak after considering the data. In contrast, the critical release rate remains uniform even after considering the experimental data in the calibration. This indicates that the Young's modulus and damage viscosity parameters can be inferred with these QoIs, but the critical release rate parameter is not identifiable based on the data. 

    Three initial conditions were used to investigate the impact of the initial geometry on the final calibration without explicitly modeling features of the microstructure geometry as an uncertain parameter in the calibration procedure. The marginal posterior distributions for the damage viscosity have similar peak values and a high degree of overlap for all three meshes. This indicates that for the damage viscosity, the initial condition has little impact on the calibration result. 
    
    In the marginal posterior for Young's modulus, there is more separation in the distributions. The distributions for the M-50-02 and M-50-01 meshes overlap, indicating that they share many probable values. The M-50-03 mesh distribution is somewhat higher than the other two, but there is overlap between the upper tails of the distributions for M-50-02 and M-50-01 and the lower tail of the M-50-03 distribution. Also, the upper tail of the M-50-02 mesh distribution is cut off by the upper bound of the uniform prior at $300\ \si{\mega\pascal}$. If the same extended prior range for M-50-03 had been used with M-50-02, there may be more overlap between the Young's modulus marginal distributions for M-50-02 and M-50-03. Additionally, all three MAP values are within 10\% of each other, which is within the expected range for material properties of heterogeneous materials modeled as continua. Moreover, all three sets of MAP values lead to similar force-displacement data when used as the parameters in a new DNS (see Section \ref{sec:MAPconfirmation}). Overall, the calibration results for the Young's modulus change slightly with the different meshes. The impact on this parameter is likely due to how the IDOX grains are arranged in the specimen, which can alter the stiffness of the specimen or how damage accumulates, depending on the grain distribution and orientation. 

    \begin{table}[h!]
    \centering
    \begin{tabular}{lllll}
        \hline
        & $\hat{E}^{(M)}$ ($\si{\mega\pascal}$) & $\hat{G}_c^{(M)}$ ($\si{\pico\joule\per\micro\meter\squared}$) & $\hat{\xi}^{(M)}$ ($\si[inter-unit-product = \ensuremath{{}\cdot{}}]{\mega\pascal\second}$) \\
        \hline
        \hline
        M-50-02 & 295.477 & 1.214 & 0.339 \\ 
        M-50-03 & 317.839 & 3.997 & 0.294 \\ 
        M-50-01 & 286.181 & 2.417 & 0.306 \\ 
        M-50-01-Cal90 & 403.769 & 3.847 & 0.327 \\ 
        \hline
    \end{tabular}
    \caption{MAP values for the four calibration cases. These summarize the best calibrated values for the parameters. Note that the critical release rate marginal posteriors are approximately uniform; hence, all points in the marginal posterior are equally probable and can serve as the MAP value. }
    \label{tab:calibration-MAPvalues}
    \end{table}

    \subsection{Calibration with $90\ \si{\degreeCelsius}$ Data}
    \label{sec:results-90C}

\begin{figure}[!h]
        \centering
        \includegraphics[scale=1.1]{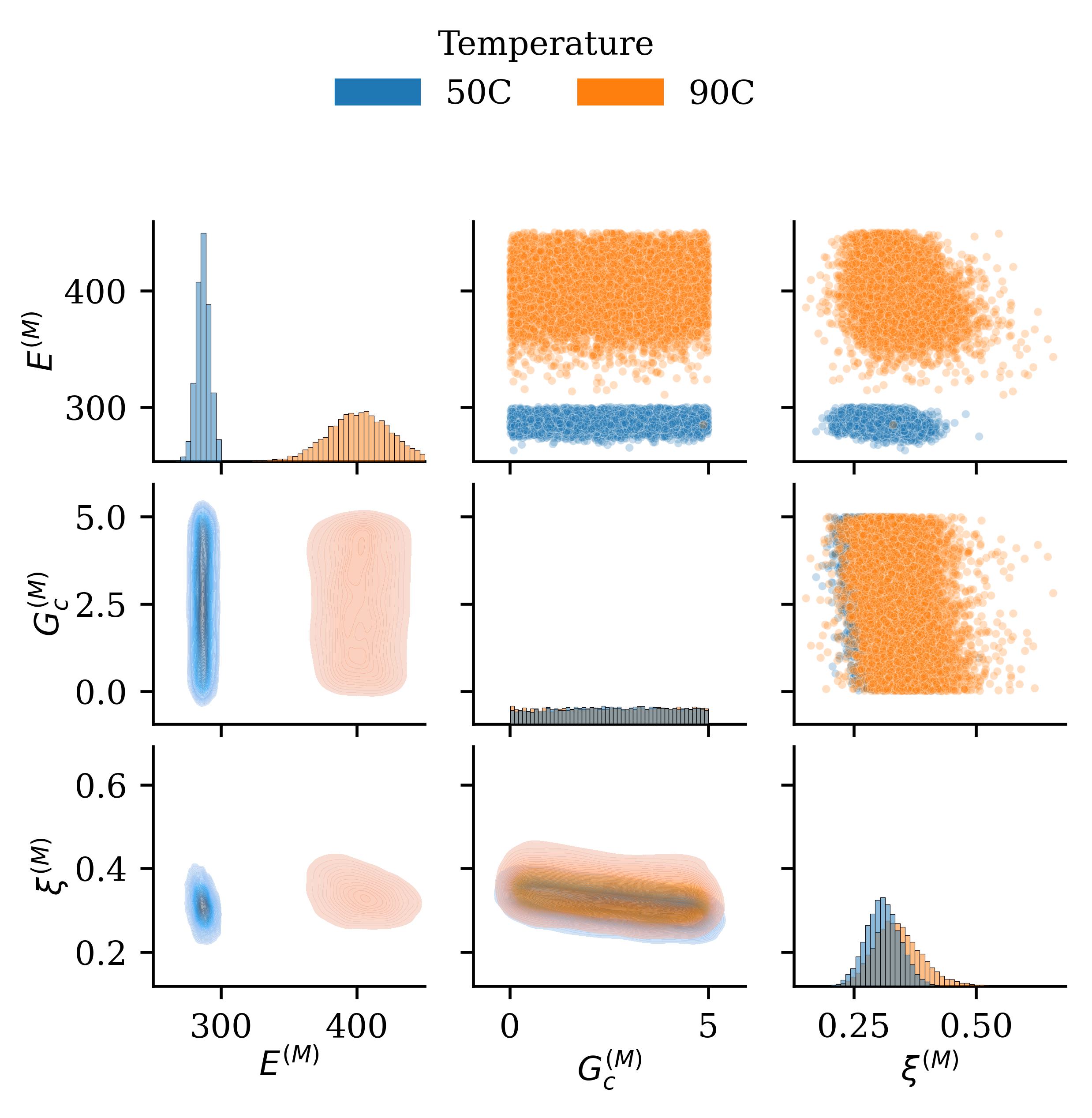}
        \caption{Posteriors for the calibrations with $50\ \si{\degreeCelsius}$ and $90\ \si{\degreeCelsius}$ data for the M-50-01 mesh. }
        \label{fig:posteriors-90C}
    \end{figure}

    The calibration procedure was also applied to the $90\ \si{\degreeCelsius}$ data. Since the initial geometry was minimally impactful for two of the three parameters in the $50\ \si{\degreeCelsius}$ calibration, only simulations with the M-50-01 mesh were used here. The posterior is shown in contrast to the $50\ \si{\degreeCelsius}$ calibration with the same mesh in Figure \ref{fig:posteriors-90C}. The MAP values are provided in Table \ref{tab:calibration-MAPvalues}. Like the $50\ \si{\degreeCelsius}$ calibration cases, the Young's modulus and damage viscosity parameters were identifiable based on the calibration data, but the critical release rate was not.  

    The peak force and initial slope of the $90\ \si{\degreeCelsius}$ experimental data are larger than those QoIs for the $50\ \si{\degreeCelsius}$ data (see Figure \ref{fig:qois-all}). This primarily impacts the calibration of the Young's modulus. The MAP value for the $90\ \si{\degreeCelsius}$ case is approximately 140\% higher than the $50\ \si{\degreeCelsius}$ one. Additionally, the distributions are distinct, and the $90\ \si{\degreeCelsius}$ distribution is much wider than the $50\ \si{\degreeCelsius}$ distribution. This is likely due to the $90\ \si{\degreeCelsius}$ experimental data having more variation in the QoIs, particularly the initial slope, and only having four data sets compared to the eight $50\ \si{\degreeCelsius}$ data sets. Interestingly, the damage viscosity marginal posterior distribution and MAP values for both temperatures are very similar. This suggests that the higher manufacturing temperature primarily impacts the stiffness of the specimens. 

\subsection{Confirming Calibration Results}
\label{sec:MAPconfirmation}

\begin{figure}[!h]
        \centering
        \subfloat[$50\ \si{\degreeCelsius}$]{\includegraphics[width=0.48\linewidth]{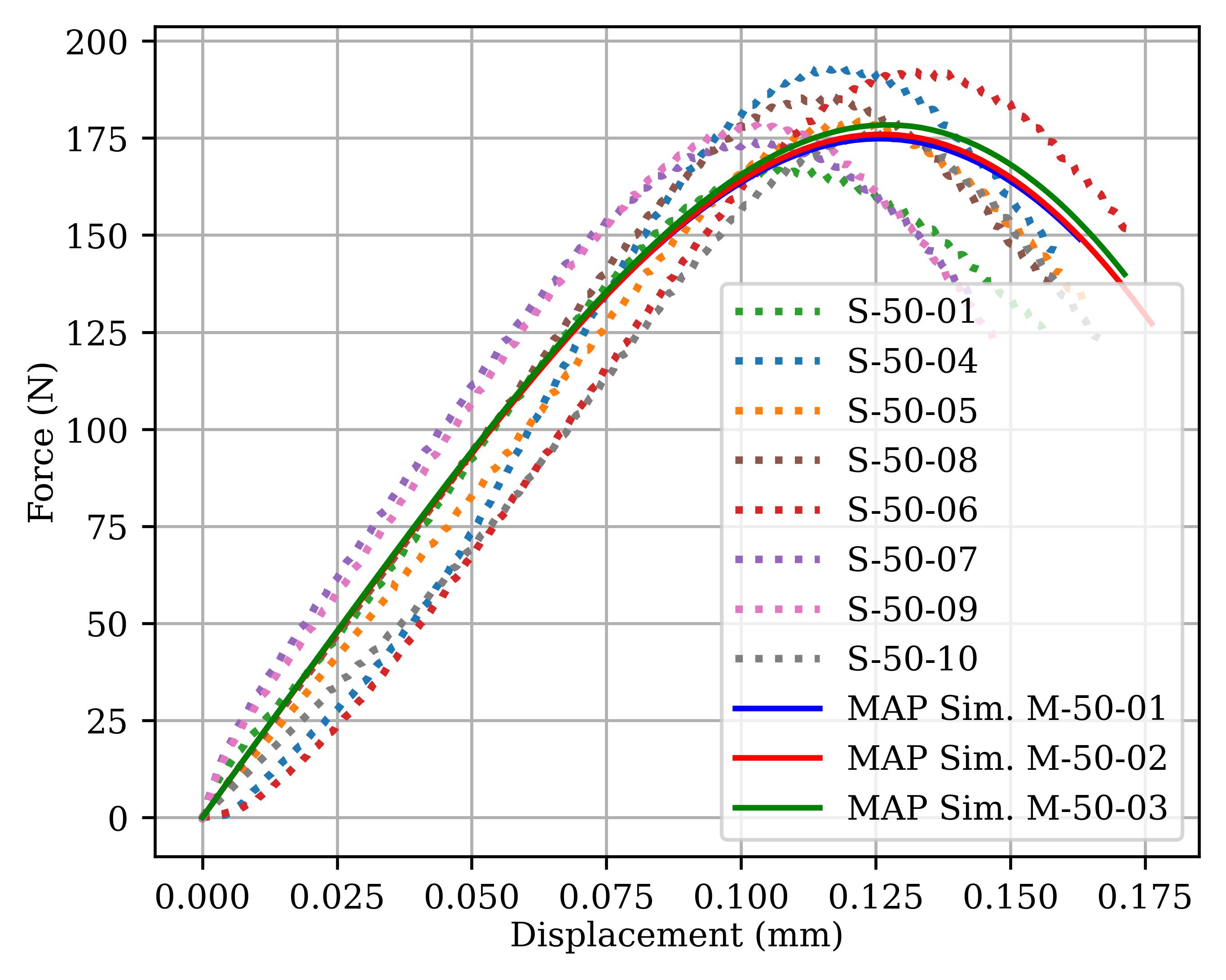}}
        \hfill
        \subfloat[$90\ \si{\degreeCelsius}$]{\includegraphics[width=0.491\linewidth]{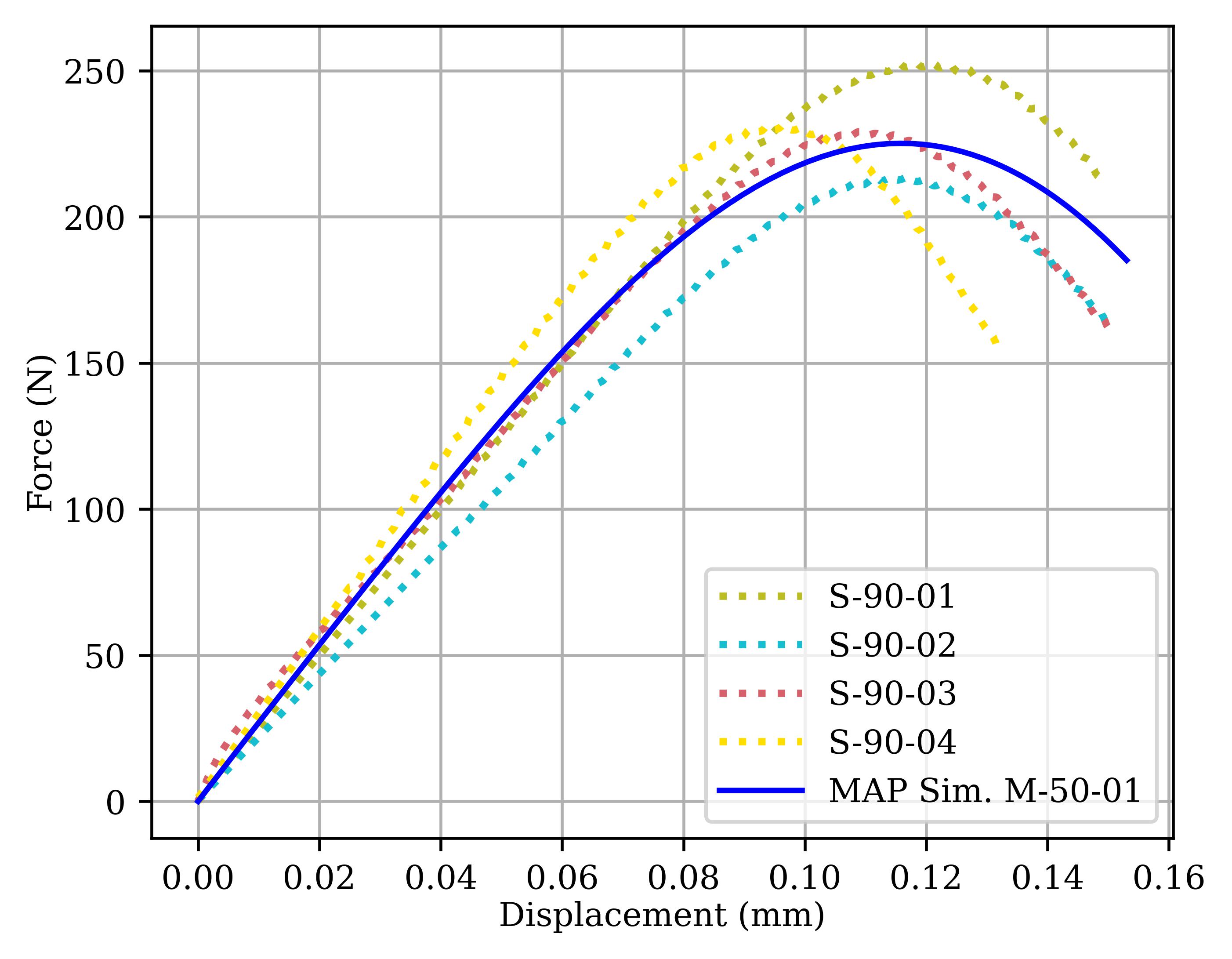}}    
        \caption{Experimental (dashed lines) and MAP simulation (solid lines) force-displacement data.}
        \label{fig:MAP-force-dispalacement}
    \end{figure}

    To confirm that the calibrated parameters result in models that replicate the behavior of the experimental data used in the calibration, new Ratel simulations were performed using the MAP values (Table \ref{tab:calibration-MAPvalues}) and the mesh used in the calibration simulations. The force-displacement data for the MAP simulations is shown with the experimental data in Figure \ref{fig:MAP-force-dispalacement}. Qualitatively, the MAP simulations approximate the average behavior of the experimental data used in the calibrations. In the $50\ \si{\degreeCelsius}$ case, all three MAP simulations lie approximately on top of each other, demonstrating that the posterior variation in the parameters leads to similar macro-scale behavior. 

\begin{figure}[!h]
        \centering
        \subfloat[$50\ \si{\degreeCelsius}$]{\includegraphics[width=0.98\linewidth]{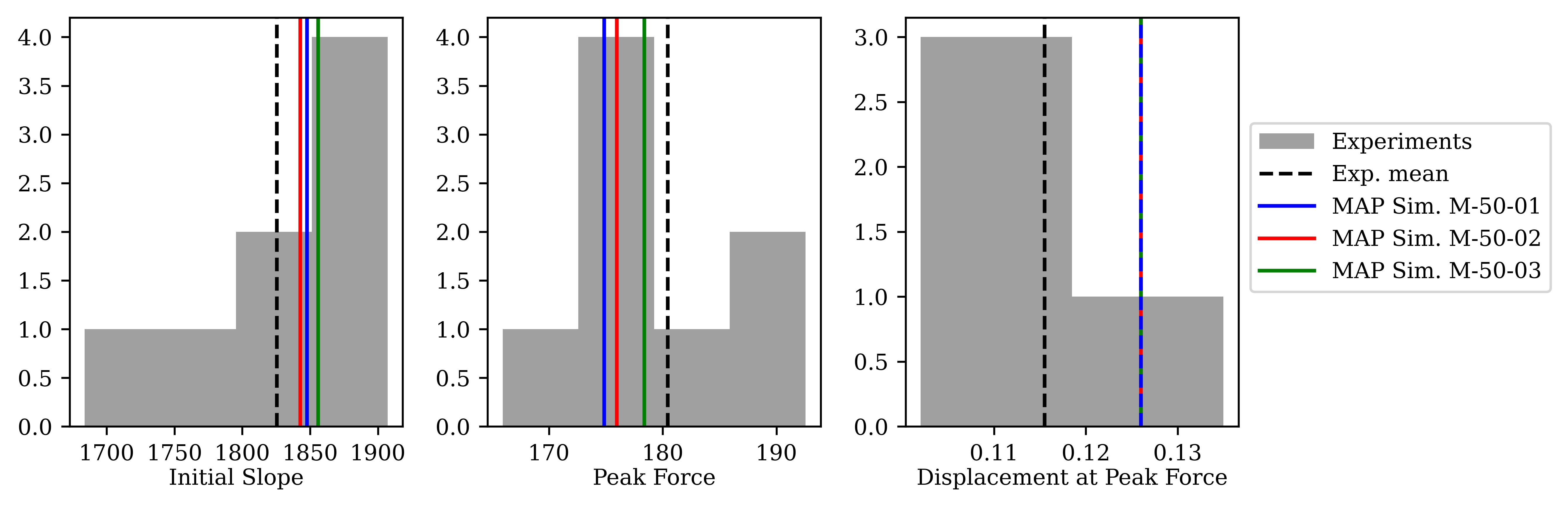}}\\
        \subfloat[$90\ \si{\degreeCelsius}$]{\includegraphics[width=0.98\linewidth]{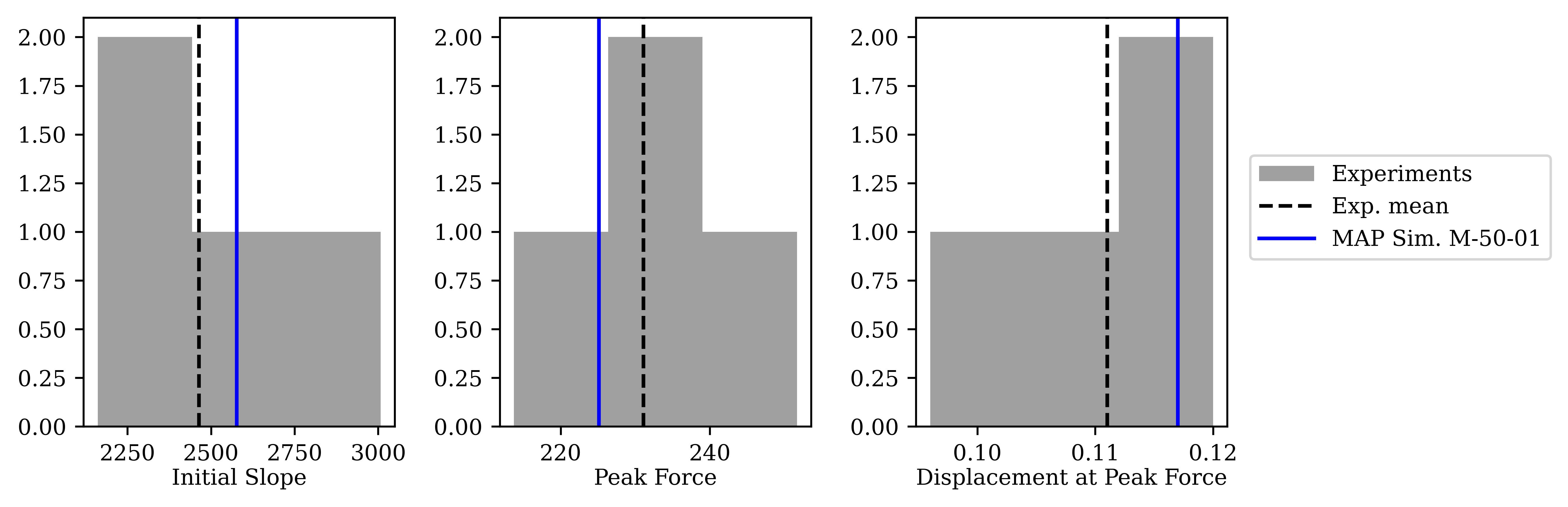}}
        \caption{Experimental (gray histogram), average experimental (dashed line), and MAP simulation (solid lines) QoIs computed from the force-displacement data shown in Figure \ref{fig:MAP-force-dispalacement}.}
        \label{fig:MAP-qois}
    \end{figure}

    The same QoIs used for calibration were also calculated for the MAP simulations. The experimental and MAP QoIs are plotted in Figure \ref{fig:MAP-qois}. The QoIs from the MAP simulations fall within the range of experimental QoIs, and the MAP QoIs are close to the experimental averages. Additionally, the percent error between the MAP simulation QoIs and the mean experimental data QoIs is reported in Table \ref{tab:MAPpercenterror}, and all of the MAP QoIs are within 10\% of the mean experimental QoIs. Thus, the calibrated models can reproduce the experimental data used for calibration within a reasonable 10\% margin. 

    \begin{table}[h!]
    \centering
    \begin{tabular}{lllll}
        \hline
        &  Initial slope & Peak force & Displacement at peak force \\
        \hline
        \hline
        M-50-02 & 0.946 & 2.478 & 9.091 \\ 
        M-50-03 & 1.674 & 1.141 & 9.091 \\ 
        M-50-01 & 1.225 & 3.101 & 9.091 \\ 
        M-50-01-Cal90 & 4.603 & 2.582 & 5.405 \\ 
        \hline
    \end{tabular}
    \caption{Percent error between the mean experimental QoIs and QoIs computed from the MAP simulation.}
    \label{tab:MAPpercenterror}
    \end{table}


\section{Conclusions} 
\label{sec:conclusions}

    This work utilized a Bayesian inference framework to calibrate three material parameters for the matrix with uncertainty quantification using data from quasi-static unconfined compression experiments and DNS. The DNS were run with three initial conditions constructed from CT scans on experimental specimens, and it was found that the initial geometry had a negligible impact on the calibration results for the damage viscosity and critical release rate parameters. The Young's modulus calibration results showed more variation with the change in initial mesh, likely due to the initial locations of the IDOX grains impacting the initial stiffness and how damage propagates through the specimen. Including information about the initial condition geometry as a feature of the calibration data is currently an open question and could be an avenue for future work to quantify how much of the variation in the parameters is due to uncertainty in the material response compared to geometric effects. Experimental data from specimens manufactured at two temperatures were also considered. The higher manufacturing temperature led to a higher Young's modulus for the matrix material. In all calibration cases, the critical release rate parameter was unidentifiable with the data and QoIs. Overall, this work demonstrates a workflow for parameter calibration with uncertainty quantification using experimental calibration data and advanced DNS. 

    Since the framework and workflow are flexible, they can be adjusted to calibrate additional parameters or different models, include new experimental data, and utilize different QoIs. For example, with these materials, it is also of interest to model how damage accumulation impacts the behavior after reaching peak force. Thus, future work could extend this calibration to post-peak behavior utilizing newer viscoelasticity and damage models, and additional QoIs in the post-peak regime. Additionally, challenges with meshing the initial geometry from CT data could be addressed by using a material point method approach rather than finite elements. In all, updates to the simulations and subsequent calibrations would result in well-calibrated constitutive models that could be used in downstream simulation efforts for mock high explosives to further investigate damage patterns and the mechanical behavior with respect to the microstructure.


\section*{Acknowledgments}

    This work was supported by the Department of Energy, National Nuclear Security Administration, Predictive Science Academic Alliance Program (PSAAP), Award Number DE-NA0003962. The Tescan S8252G scanning electron microscope used to perform the SEM imaging presented in this work was acquired with funds from the National Science Foundation, Award Number DMR-1828454). H. Lu also acknowledges the Louis A. Beecherl Jr. Chair for additional support. NEP and AJC acknowledge support from the U.S. Department of Energy through the Los Alamos National Laboratory during the preparation of this manuscript. Los Alamos National Laboratory is operated by Triad National Security, LLC, for the National Nuclear Security Administration of U.S. Department of Energy (Contract No. 89233218CNA000001).

    The authors would like to thank the other members of the PSAAP 3 Multi-disciplinary Simulation Center for Micromorphic Multiphysics Porous and Particulate Materials Simulations Within Exascale Computing Workflows and the Ratel development team \cite{ratel-user-manual}. 
    

\section*{Data Statement}

The raw and processed data required to reproduce the above findings cannot be shared at this time, as the data also forms part of an ongoing study.


\appendix

\section{Constitutive Model}
\label{sec:app-materialmodel}

    The adopted constitutive model formulation, which couples hyperelasticity with the phase-field modeling of fracture via a monolithic scheme, is described in detail in \cite{ratel-user-manual}. Here, we include the expressions that introduce the relevant quantities. The strain energy $\psi$ is defined as a function of the Eulerian logarithmic strain tensor $\textbf{e}^e$ and fourth-order elasticity tensor $\mathsf C$,
    \begin{equation}
        \psi(\boldsymbol{e}^e) \equiv \frac{1}{2}\textbf{e}^e:\mathsf C:\textbf{e}^e. 
    \end{equation} 

    The logarithmic strain can be defined via the elastic left stretch tensor $\textbf{v}^e$ or the elastic left Cauchy-Green strain tensor $\boldsymbol{b}^e=(\textbf{v}^e)^2=\boldsymbol{F}^e(\boldsymbol{F}^e)^T$
    \begin{equation}
        \textbf{e}^e \equiv \log{\textbf{v}^e}=\frac{1}{2}\log\boldsymbol{b}^e = \frac{1}{2}\sum_{i=1}^3\log(\lambda^b_i)\hat{\boldsymbol{n}}_i \otimes \hat{\boldsymbol{n}}_i,
    \end{equation}
    with eigenvalues $\lambda^b_i=\lambda^2_i$ and eigenvectors $\hat{\boldsymbol{n}}_i$. 

    The symmetric Kirchoff stress tensor $\boldsymbol{\tau}$ can be written, 
    \begin{equation}
        \boldsymbol{\tau} = \frac{\partial \psi}{\partial \textbf{e}^e} = \mathsf C:\textbf{e}^e = 2\mu \textbf{e}_d^e + K tr(\textbf{e}^e) \boldsymbol{I},
    \end{equation}

    where $\textbf{e}_d^e$ is the deviatoric component of $\textbf{e}^e$, $\mu = E/2(1+\nu)$, and $K=E/3(1-2\nu)$. 

    For the damage model, a phase field model for fracture is used, where a variational formulation for brittle fracture is employed with the crack surface density function 
    \begin{equation}
        \gamma(\phi, \nabla\phi) = G_c \frac{1}{c_0 l_0} [\alpha(\phi) + l_0^2 \vert\nabla\phi\vert^2], 
    \end{equation}
    where $0\le \phi \le 1$ is the damage phase field, $c_0 = 4\int^{1}_{0}\sqrt{\alpha(\phi)}d\phi$ is a scaling factor, $l_0$ is the length scale that controls the ``thickness'' of the smeared crack in the phase-field modeling of fracture, $\alpha(\phi)$ is the geometric crack function, and $G_c$ is the critical energy release rate. The value of $G_c$ can be derived from the critical stress ($\sigma_c$). For the Ambrosio–Tortorelli-2 (AT2) model is \cite{TanneModeling2018}
    \begin{equation}
        \sigma_c = \frac{3}{16}\sqrt{\frac{3G_c E'}{l_0}},
    \end{equation}
    where $E'=E/(1-\nu^2)$ in plane strain. 
    The crack surface energy can be computed by integrating over the domain $\Omega$,
    \begin{equation}
        \Psi_{\Gamma}=\int_{\Gamma}G_c dA \approx \int_{\Omega}G_c \gamma(\phi;\nabla\phi)dV. 
    \end{equation}

    In this formulation, the strain energy stored in the solid can be expressed as the elastic strain energy $\psi$ multiplied by a degradation function $g(\phi) = (1-\eta)(1-\phi)^2 + \eta$ where $\eta \ll 1$ is the residual stiffness and prevents numerical instability. Additionally, Ratel implements a viscous regularization to prevent convergence loss by complementing the damage residual with a damage viscosity ($\xi$) term.

\bibliography{ref}

\end{document}